\documentclass[a4paper,UKenglish,cleveref, autoref]{lipics-v2021}

\title{Matching and Edge Cover in Temporal Graphs} 

\author{Lapo Cioni}{Università degli Studi di Firenze, Italy}{lapo.cioni@unifi.it}{https://orcid.org/0000-0002-4605-8473}{}

\author{Riccardo Dondi}{Università degli Studi di Bergamo, Italy}{riccardo.dondi@unibg.it}{https://orcid.org/0000-0002-6124-2965}{}

\author{Andrea Marino}{Università degli Studi di Firenze, Italy}{andrea.marino@unifi.it}{}{}

\author{Jason Schoeters}{Università degli Studi di Firenze, Italy}{jason.schoeters@unifi.it}{https://orcid.org/0000-0001-7257-5426}{}

\author{Ana Silva}{Universidade Federal do Ceará, Brasil}{anasilva@mat.ufc.br}{}{}

\authorrunning{Cioni L., Dondi R., Marino A., Schoeters J. Silva A.}



\acknowledgements{This paper was partially funded
by Italian PNRR CN4 Centro Nazionale per la Mobilità Sostenibile, NextGeneration EU - CUP,  B13C22001000001. MUR of Italy, under PRIN Project n. 2022ME9Z78 - NextGRAAL: Next-generation algorithms for constrained GRAph visuALization, PRIN PNRR Project n. P2022NZPJA - DLT-FRUIT: A user centered framework for facilitating DLTs FRUITion.}

\usepackage{amsmath,amssymb,amsthm}
\usepackage{packages/slashbox}
\usepackage[normalem]{ulem}

\usepackage{float}
\usepackage{complexity}
\usepackage{makecell}
\usepackage[table]{xcolor}

\usepackage[ruled]{algorithm2e}
\SetKwRepeat{Do}{do}{until}

\graphicspath{{./tikzdrawings}{.}}

\newcommand{\myparagraph}[1]{\noindent\textbf{#1}}

\newcommand{\threeSAT}[1]{{\textnormal{\textsc{3-SAT}#1}}}
\newcommand{\TMatching}[1]{{\textnormal{\textsc{Temporal Matching}#1}}}
\newcommand{\TEdgeCover}[1]{{\textnormal{\textsc{Temporal Edge Cover}#1}}}
\newcommand{\SetCover}[1]{{\textnormal{\textsc{Min Set Cover}#1}}}
\newcommand{\PSets}[1]{{\textnormal{\textsc{Packing Sets}#1}}}

\hideLIPIcs
\nolinenumbers

\begin{document}

\maketitle

\begin{abstract}
Temporal graphs are a special class of graphs for which a temporal component is added to edges, that is, each edge possesses a set of times at which it is available and can be traversed. Many classical problems on graphs can be translated to temporal graphs, and the results may differ.

In this paper, we define the \TEdgeCover{} and \TMatching{} problems and show that they are NP-complete even when fixing the lifetime or when the underlying graph is a tree. We then describe two FPT algorithms, with parameters lifetime and treewidth, that solve the two problems. 
We also find lower bounds for the approximation of the two problems and 
give two approximation algorithms which match these bounds. 
Finally, we discuss the differences between the problems in the temporal and the static framework.
\end{abstract}

\section{Introduction}


A temporal graph is a graph where the edges are available only at prescribed moments. More formally, a \emph{temporal graph} with \emph{lifetime} $\tau$ is a pair ${\cal G} = (G,\lambda)$ where $G$ is a graph (called the \emph{underlying graph}) and $\lambda$ is the \emph{time labelling} that assigns to each edge a finite non-empty subset of $[\tau]$.
Alternatively, a temporal graph can be seen as a finite sequence of spanning subgraphs of $G$ called \emph{snapshots}. A \emph{temporal vertex} is an occurrence of a vertex in time, i.e. an element of $V(G)\times[\tau]$, and a \emph{temporal edge} is an occurrence of an edge in time, i.e. $(e,t)$ with $e\in E(G)$ and $t\in \lambda(e)$.
They appear in the literature under many distinct names (temporal networks~\cite{Holme.15}, edge-scheduled networks~\cite{B.96}, dynamic networks~\cite{XFJ.03}, time-varying graphs~\cite{CFQS.12}, stream graphs, link streams~\cite{LVM.18}, etc). We refer the reader to~\cite{Holme.15} for a plethora of applications.
In the recent years, many papers have focused on studying how well-known problems in static graph theory translate into the temporal setting. In this paper we focus on edge covering and matching problems.

A \emph{matching}\footnote{The definitions for matching and edge cover, as well as their relationship, can be found in most graph theory books. We refer to~\cite{west2001introduction}.} is a set of edges such that no two edges share a common vertex. 
An \emph{edge cover} is a set of edges ensuring that every vertex in the graph is incident to at least one edge in the set. 
The \emph{maximum matching problem} seeks to find a matching of the largest possible size, while the \emph{minimum edge cover problem} aims to determine the smallest edge cover\footnote{We assume that the graph \( G \) has no isolated vertices.}. These are fundamental problems in graph theory, known to be dual to each other and solvable in polynomial time.  
To illustrate their duality, consider a maximum matching \( M \) in a graph \( G \). A minimum edge cover \( S \) of size \( |V(G)|-|M|\) can be obtained from  \( M\) by greedily adding edges until all vertices in \( G \) are covered.
Applying similar combinatorial reasoning, one can obtain a maximum matching from a minimum edge cover, bringing us to the equality $\alpha'(G)+\beta'(G)=|V(G)|$~\cite{gallai1959uber}, where \( \alpha'(G) \) is the size of a maximum matching and \( \beta'(G) \) is the size of a minimum edge cover. This is known as Gallai's Theorem.

The above concepts naturally extend to temporal graphs in multiple ways, depending on whether we aim to cover or saturate vertices versus temporal vertices, and whether we achieve this using edges or temporal edges. This distinction gives rise to four possible variations, as summarized in Tables~\ref{tab:ECvariants} and~\ref{tab:Mvariants}.  
It is straightforward to show that most of these variations reduce to solving the corresponding minimum edge cover or maximum matching problem in static graphs. 
Indeed, whenever vertices are considered, the temporal component of the edges does not play a role in the problems, and the solutions are the same as those of the corresponding static problems on the underlying graph. On the other hand, if both temporal edges and temporal vertices are considered, then the snapshots of the temporal graph are independent and can be solved as they where static graphs (the resulting graph is called static expansion of a temporal graph~\cite{staticIntro}). 
For this reason, we focus on the cases highlighted in pink. In the following, we formally define the relevant concepts. We say that a temporal vertex $(v,t)$ is \emph{isolated} if $t\notin \lambda(uv)$ for every $u\in N(v)$ (in other words, if $v$ is isolated in snapshot $G_t$).

\begin{table}[t]
\centering
\begin{tabular}{|c|c|c|} 
\hline
\backslashbox{covered}{covered by} & \textsc{temporal} \textsc{edge} & \textsc{edge} \\ 
\hline
\hline
\textsc{temporal} \textsc{vertex} & polynomial & \cellcolor{red!25}{\makecell{\NP-complete (Theorem.~\ref{thm:mainec})}} \\
\hline
\textsc{vertex} & polynomial & polynomial \\
\hline
\end{tabular}
\caption{Temporal variations of edge cover. }
\label{tab:ECvariants}
\end{table}

\begin{table}[t]
\centering
\begin{tabular}{|c|c|c|} 
\hline
\backslashbox{not sharing}{taking} & \textsc{temporal} \textsc{edge} & \textsc{edge} \\ 
\hline
\hline
\textsc{temporal} \textsc{vertex} & polynomial & \cellcolor{red!25}{\makecell{\NP-complete (Theorem~\ref{thm:mainm})}} \\
\hline
\textsc{vertex} & polynomial & polynomial \\
\hline
\end{tabular}
\caption{Temporal variations of matching. }
\label{tab:Mvariants}
\end{table}


\begin{definition}[Temporal Edge Cover]
Given a temporal graph $\mathcal{G}=(G,\lambda)$, a \emph{temporal edge cover} of $\mathcal{G}$ is a subset $S\subseteq E(G)$ such that, for every non-isolated $(v,t)\in V(G)\times [\tau]$, there exists an edge $e\in S$ incident to $v$ such that $t\in\lambda(e)$.
\end{definition}

\begin{figure}[t]
\begin{minipage}{\linewidth}
    \centering
    \includegraphics[scale=1.5]{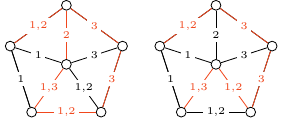}
    \caption{Two minimal temporal edge covers of a temporal graph. The one on the right has minimum cardinality.}
    \label{exEC}
\end{minipage}

\begin{minipage}{\linewidth}
    \centering
    \includegraphics[scale=1.5]{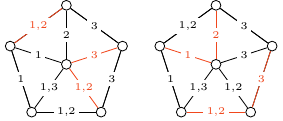}
    \caption{Two maximal temporal matchings of a temporal graph. The one on the right has maximum cardinality.}
    \label{exM}
\end{minipage}
\end{figure}

Examples of temporal edge cover are shown in \cref{exEC}. 
Observe that the temporal edge covers presented are minimal, with the one on the right having the smallest cardinality among all edge covers of that temporal graph.

\begin{definition}[Temporal Matching]
Given a temporal graph $\mathcal{G}=(G,\lambda)$, a subset $M\subseteq E(G)$ is a \emph{temporal matching} of $\mathcal{G}$ if for every $e,e'\in M$, $e\neq e'$, either $e\cap e'=\emptyset$ or $\lambda(e)\cap\lambda(e')=\emptyset$.
\end{definition}

Examples of temporal matching are shown in \cref{exM}. Observe that the temporal matchings are {maximal}, with the one on the right having maximum cardinality among all temporal matchings.

We call \TEdgeCover{} (resp. \TMatching{})  the problem of, given a temporal graph $\mathcal{G}$ and a nonnegative integer $k$, deciding whether there exists a temporal  edge cover (resp. temporal matching) of $\mathcal{G}$ of size at most (resp. at least) $k$.

\myparagraph{Our Contributions.} 
Our results are summarized in Theorems~\ref{thm:mainec} and~\ref{thm:mainm}.  
We prove that both problems are \NP-complete, even when \(\tau = 2\) or when the underlying graph is a tree.  
This implies that both problems are para-\NP-complete when parameterized by either the lifetime or the treewidth of the underlying graph.  
We then show that combining these parameters allows us to obtain \(\FPT\) algorithms.  
It is worth noting that the apparent similarity between the two problems is not due to shared proof techniques; rather, all proofs are independent.  
Finally, the problems differ in terms of approximation: while \TEdgeCover{} can be approximated within a logarithmic factor, \TMatching{} cannot.  
In particular, note that our approximation factors are asymptotically optimal. 


\begin{theorem}
\TEdgeCover{}
\begin{enumerate}
    \item is \NP-complete even if $\tau=2$;
    \item is \NP-complete even if the underlying graph is a tree;
    \item is \FPT{} parameterized by $\tau$ plus the treewidth of the underlying graph;
    \item cannot be approximated within factor $b\log\tau$ for any $b$ with $0<b<1$, unless \P=\NP;
    \item can be approximated within factor $O(\log\tau)$.
\end{enumerate}
\label{thm:mainec}
\end{theorem}

\begin{theorem}
\TMatching{}
\begin{enumerate}
    \item is \NP-complete even if $\tau=2$;
    \item is \NP-complete even if the underlying graph is a tree;
    \item is \FPT{} parameterized by $\tau$ plus the treewidth of the underlying graph;
    \item cannot be approximated within factor 
    $\tau^{1-\varepsilon}$, for any $\varepsilon > 0$.
    \item can be approximated within factor $\tau$.
\end{enumerate}
\label{thm:mainm}
\end{theorem}


As previously noted, despite the apparent similarity, the proofs of Theorems~\ref{thm:mainec} and~\ref{thm:mainm} are fundamentally different. This independence arises from the fact that the size of a minimum temporal edge cover is unrelated to the size of a maximum temporal matching, unlike the case of static graphs. 
In fact, in Section~\ref{sec:rel} we prove a stronger result, namely that having a minimum temporal edge cover does not facilitate the computation of a maximum temporal matching, and vice versa. More specifically, we show that, given a temporal matching of maximum cardinality, finding a minimum temporal edge cover remains \NP-complete. Likewise, given a minimum temporal edge cover, finding a maximum temporal matching is also \NP-complete. Observe that this implies that a temporal version of Gallai's Theorem cannot hold unless \(\P=\NP\). 

\myparagraph{Related Works.}
Many variations of the temporal matching problem have been explored in the literature.  
The first definition of a temporal matching appears in~\cite{IntroTM}, where it is defined as a set of temporal edges \(\{(e_1,t_1),\dots,(e_q,t_q)\}\) such that \(\{e_1,\dots,e_q\}\) forms a matching in the underlying static graph, and all timestamps are distinct.  
This constraint can be quite restrictive, as it permits selecting at most one edge per snapshot.  

A relaxation of this constraint was introduced in~\cite{deltaMatching} with the concept of a \(\Delta\)-temporal matching. In this variation, temporal edges incident to the same vertex must have timestamps that differ by at least \(\Delta\). This concept arises from the idea of analyzing the graph through temporal windows of size \(\Delta\), which led to the definition of several \(\Delta\)-related problems, summarized in~\cite{deltaProblems}. In the latter work, they also introduce the notion of a \(\Delta\)-edge cover, leaving open the related problem. 

A closely related concept is that of a \(\gamma\)-matching in a link stream, introduced in~\cite{gammaMatching}, where \(\gamma\) is a fixed positive integer. Using our terminology, this corresponds to a set of temporal edges \(\{(e_1,t_1),\dots,(e_q,t_q)\}\) such that \(\{t_i,\dots,t_i+\gamma-1\} \subseteq \lambda(e_i)\) for each \(i \in [q]\), and whenever \(|t_i - t_j| < \gamma\), then \(e_i \cap e_j = \emptyset\). Observe that this is a special case of \(\Delta\)-temporal matching.

\section{Preliminaries}

A (undirected, loopless) graph $G$ is an ordered pair $(V,E)$, where $V$ is a finite set and $E\subseteq \{\{u,v\}\mid u,v\in V, u\neq v\}$. The elements of $V$ are called \emph{vertices} and the elements of $E$ are called edges. Sometimes we use $V(G)$ and $E(G)$ to refer to the set of vertices and edges of $G$, respectively.
Also, for simplicity, we write the elements of $E(G)$ as $uv$ instead of $\{u,v\}$, while still using the notation $u\in uv$.
Given $v\in V(G)$, let $\delta_{G}(v)=\{e\in E(G)\mid v\in e\}$ be the set of edges incident to $v$ in $G$.
%
Given a graph $G$, a positive integer $\tau$ and a function $\lambda: E(G) \rightarrow \mathcal{P} ([\tau])$, with $\mathcal{P}([\tau])$ being the power set of $\{1,\dots,\tau\}$, such that each edge is assigned a finite non-empty subset of $[\tau]$. 
Then $\mathcal{G} = (G,\lambda)$ is a \emph{temporal graph} with \emph{lifetime} $\tau$.
%
We can see the vertices and edges of $\mathcal{G}$ in two ways. One is to see them as just the vertices and edges of $G$. The other is to add a temporal component to them. In this way, we have \emph{temporal vertices} in the form $(v,i)\in V(G)\times [\tau]$, and \emph{temporal edges} in the form $(e,j)$ with $e\in E(G)$ and $j\in\lambda(e)$.
We recall some NP-complete problems that we use in the reductions of this paper.

\threeSAT{(2,2)}: given an input boolean formula $F$ in conjunctive normal form, where each clause has three literals and each variable appears four times, of which exactly two times is negated, decide whether $F$ is satisfiable and, if so, give an assignment that satisfies it.

\SetCover{}: given a pair $(U,\mathcal{S})$ and a nonnegative integer $k$, where $U=[n]$ for some $n$ and $\mathcal{S}=\{S_1, \dots, S_m\}$ is a collection of subsets of $U$, determine (if it exists) a subcollection of at most $k$ subsets $S_{i_1}$, \dots $S_{i_k}$ such that $U\subset\bigcup_{j= 0}^{k} S_{i_j}$.

\PSets{}: given a collection of sets $\mathcal{S}=\{S_1, \dots, S_m\}$ and a nonnegative integer $k$, determine (if it exists) a subcollection of at least $k$ pairwise disjoint sets in $\mathcal{S}$.

Finally, we recall the definition of \emph{nice tree decomposition}, that we use for the \FPT{} algorithms.

A \emph{tree decomposition} of a graph \( G \) is a pair \( (T, \{X_t\}_{t \in V(T)}) \), where \( T\) is a tree and \( \{X_t\}_{t \in V(T)} \) is a collection of subsets of \( V(G) \) (called bags), such that the following three conditions hold:

\begin{enumerate}
   \item Every vertex of \( G \) appears in at least one bag:  
   \[
   \bigcup_{t \in V(T)} X_t = V(G).
   \]

	\item 
   For every edge \( (u, v) \in E(G) \), there exists a bag \( X_t \) such that both \( u \) and \( v \) are in \( X_t \):  
   \[
   \forall (u, v) \in E(G), \exists t \in V(T) \text{ such that } \{u, v\} \subseteq X_t.
   \]

	\item
   For every vertex \( v \in V(G) \), the set of nodes \( \{t \in V(T) \mid v \in X_t\} \) forms a subtree of \( T \).
\end{enumerate}

The \emph{width} of a tree decomposition is defined as \( \max_{t \in V(T)} |X_t| - 1 \), i.e., the size of the largest bag minus one. The \emph{treewidth} of a graph \( G \) is the minimum width over all possible tree decompositions of \( G \).

A tree decomposition \( (T, \{X_t\}_{t \in V(T)}) \) of \( G \) is a \emph{nice tree decomposition} if:
\begin{enumerate}
	\item \( T \) is a rooted tree (call $r$ its root), and each node \( t \in V(T) \) is one of the following types:
	\begin{itemize}
  		\item \emph{Leaf node}: \( t \) is a leaf of \( T \), and \( X_t = \emptyset \).
   		\item \emph{Introduce node}: \( t \) has exactly one child \( t' \), and \( X_t = X_{t'} \cup \{v\} \) for some \( v \notin X_{t'} \). We say that $t$ \emph{introduces $v$}.
   		\item \emph{Forget node}: \( t \) has exactly one child \( t' \), and \( X_t = X_{t'} \setminus \{v\} \) for some \( v \in X_{t'} \). We say that $t$ \emph{forgets $v$}.
   		\item \emph{Join node}: \( t \) has exactly two children \( t_1 \) and \( t_2 \), and \( X_t = X_{t_1} = X_{t_2} \). 
	\end{itemize}

	\item \( B_r = \emptyset \).
\end{enumerate}

It is largely known that a nice tree decomposition can be obtained from a tree decomposition without increasing the width. We refer the reader to~\cite{cygan2015parameterized} for a very good introduction about how to obtain algorithms that run in \FPT{} time when parameterized by the treewidth.

\section{Hardness and Tractability of \TEdgeCover{}}
\label{sectionEC}

In this section, we study the complexity of \TEdgeCover{}. Specifically, we show that \TEdgeCover{} is NP-complete when the lifetime $\tau$ of graph is 2, and then we show that it is NP-complete even when the underlying graph is a tree.
This suggests that both the lifetime $\tau$ and the treewidth $w$ of a graph play an important role in the complexity of \TEdgeCover{}. Indeed, we describe an \FPT{} algorithm in $\tau$ and $w$ which solves the problem.

\subsection{Hardness for $\tau=2$}
\label{subsec:hard2}

We prove that \TEdgeCover{} restricted to $\tau=2$ is NP-complete by giving a reduction from \threeSAT{(2,2)}. 
We first describe some (non temporal) graphs needed by our reduction, that have some edges marked (note that the marking is not the time labelling $\lambda$).

\begin{definition}
\label{Lidef}
Let $i$ be a positive integer. We define the graph $L_i=(V_i,E_i)$ to be a cycle with $10$ edges such that
(1) two edges of $E_i$ are marked $i$ and two edges of $E_i$ are marked $-i$ and
(2) there is one unmarked edge between edges with the same marking, and two unmarked edges between edges of opposite marking.
\end{definition}

\begin{figure}[t]
\begin{minipage}{0.45\linewidth}
    \centering
    \includegraphics[scale=0.8]{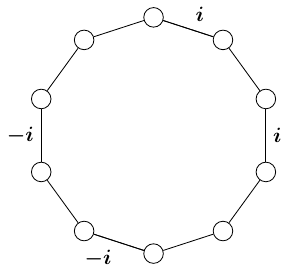}
    \caption{Graph $L_i$.}
    \label{Liimage}
\end{minipage}
\hfill
\begin{minipage}{0.45\linewidth}
    \centering
    \includegraphics[scale=0.8]{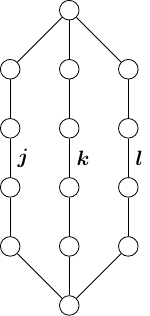}
    \caption{Graph $T_{j,k,l}$.}
    \label{Tjkl}
\end{minipage}
\end{figure}

Graph $L_i$ is shown in \cref{Liimage}. Note that, since it has ten vertices and ten edges, its vertices can be covered using five edges in two ways, denoted by $E'_{i,1}$ and $E'_{i,2}$: 

\begin{itemize}

\item $E'_{i,1}$ contains both edges marked by $i$ and no edge marked by $-i$

\item $E'_{i,2}$ contains both edges marked by $-i$ and no edge marked by $i$

\end{itemize}


Given three integers $j$, $k$, $l$ we define the graph $T_{j,k,l}$, with edges marked $j$, $k$, $l$ as in \cref{Tjkl}.

We use the graphs $L_i$ and $T_{i,j,k}$ to define an instance of \TEdgeCover{} with lifetime $2$ corresponding to an instance of \threeSAT{(2,2)}. 
Consider an instance $F$ of \threeSAT{(2,2)} consisting of clauses $C_1$, \dots, $C_m$ over $n$ variables $x_1$, \dots, $x_n$.
Recall that each $C_j$, $j \in [m]$ has three literals and each variable $x_i$, $i \in [n]$, appears in exactly two clauses as a positive literal and in exactly two clauses as a negative literal.
We construct a corresponding temporal graph $\mathcal{G}$, with lifetime $\tau=2$, associated with $F$ as follows:

\begin{itemize}

\item At time $1$, $\mathcal{G}$ is defined as a graph $G_1$ that contains, for each variable $x_i$, $i \in [n]$, a cycle $L_i$, as defined in \cref{Lidef}. Note that these cycles are all vertex disjoint.

\item At time $2$, $\mathcal{G}$ is defined as a graph $G_2$ that contains, for each clause $C_p$, $p \in [m]$, over variables $x_i$ ,$x_j$, $x_k$, with $i,j,k\in [n]$, a graph $T^p_{j,k,l}$ isomorphic to $T_{j,k,l}$.
The marked edges of $T^p_{j,k,l}$ are defined as follows. 
First, $T^p_{j,k,l}$ shares marked edges with $L_q$, $q \in \{j,k,l\}$, in $G_1$.
For each $q \in \{j,k,l\}$, if $x_q$ is positive in $C_p$, then $T^p_{j,k,l}$ and $L_q$ share an edge marked $q$, if $x_q$ is negative in $C_p$, then $T^p_{j,k,l}$ and $L_q$ share an edge marked $-q$.
Note that we define a one-to-one correspondence between the marked edges of graphs $T^q_{j,k,l}$ and of the graphs $L_i$, since each $L_i$ has two edges marked $i$ and two edges marked $-i$, and a formula in \threeSAT{(2,2)} has precisely two positive occurrences of each variable $x_i$ and two occurrences of its negation. Thus, two distinct edges of $L_i$ with the same marking corresponds to two distinct edges of some $T^p_{i,j,k}$, $T^r_{i,j',k'}$.
\end{itemize}

The resulting temporal graph can be constructed in polynomial starting from an instance $F$ of \threeSAT{(2,2)}.
Using this reduction, we can prove that $F$ is satisfiable if and only if there exists an edge cover of $\mathcal{G}$ having at most $5n+6m$ edges. The idea behind the proof is that each $L_i$ can be covered with 5 edges by $E'_{i,1}$ or $E'_{i,2}$, while each $T^p_{j,k,l}$ must be covered using at least $6$ non marked edges, with $6$ being achieved only if at least one marked edge is part of the covering. Depending on which $E'_{i,a}$ is used for the covering, true or false is assigned to the corresponding variable $x_i$.

This is formalized in the following lemmas.

\begin{lemma}
\label{no6}
The vertices of each graph $T^p_{j,k,l}$, with $p \in [m]$ and $j,k,l \in [n]$, cannot be covered using at most $6$ edges. 
\end{lemma}
\begin{proof}
Since $T^p_{j,k,l}$ has $14$ vertices, and an edge covers $2$ vertices, then $6$ edges (or less) cover at most $12$ vertices.
\end{proof}

\begin{figure}[ht]
\begin{minipage}{0.30\linewidth}
    \centering
    \includegraphics[scale=0.6]{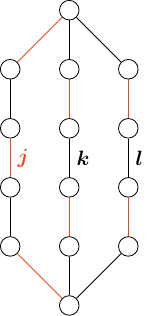}
\end{minipage}
\hfill
\begin{minipage}{0.30\linewidth}
    \centering
    \includegraphics[scale=0.6]{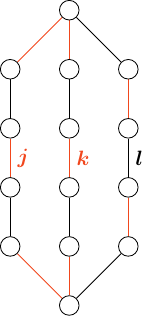}
\end{minipage}
\hfill
\begin{minipage}{0.30\linewidth}
    \centering
    \includegraphics[scale=0.6]{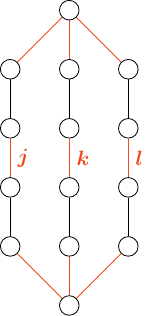}
\end{minipage}
\caption{Three ways to cover the vertices of $T_{j,k,l}$ with six edges in addition to those marked $j$, $k$, $l$.}
\label{6coverImage}
\end{figure}

\begin{lemma}
\label{6coverLemma}
Consider a graph $T^p_{j,k,l}$, with $p \in [m]$ and $j,k,l \in [n]$, and let $A^p_{i,j,k}$ be a nonempty subset of the marked edges of $T^p_{j,k,l}$. Then the temporal vertices of $T^p_{j,k,l}$:
(1) can be covered by $A^p_{j,k,l} \cup B^p_{j,k,l}$ where $B^p_{j,k,l}$ is a set of exactly six unmarked edges of $T^p_{j,k,l}$;
(2) cannot be covered by  $A^p_{j,k,l} \cup D^p_{j,k,l}$ where $D^p_{j,k,l}$ is a set of at most five unmarked edges of $T^p_{j,k,l}$.
\end{lemma}
\begin{proof}
(1) Without loss of generality, we can assume that $A^p_{j,k,l}$ contains either an edge marked $j$, two edges marked by $j,k$ or three edges marked by $j,k,l$. Then \cref{6coverImage} shows how to cover the vertices in the desired way. 

(2) Notice that there are six vertices adjacent to the top vertex of $T^p_{j,k,l}$ (see \cref{6coverImage}); these vertices require three edges to be covered. Similarly, consider the vertices adjacent to the bottom vertex of $T^p_{j,k,l}$ (see \cref{6coverImage}). These vertices require three edges to be covered.
\end{proof}

\begin{lemma}
\label{6coverLemmab}
Consider a graph $T^p_{j,k,l}$, with $p \in [m]$ and $j,k,l \in [n]$. 
Then a set $D^p_{j,k,l}$ that contains no marked edges and at most six unmarked edges of $T^p_{j,k,l}$ cannot cover every temporal vertex of $T^p_{j,k,l}$.
\end{lemma}
\begin{proof}
Assume that $D^p_{j,k,l}$ contains no marked edge and at most six unmarked edges of $T^p_{j,k,l}$. 
If $D^p_{j,k,l}$ covers every temporal vertex of $T^p_{j,k,l}$, then it must cover the six endpoints of the marked edges with six distinct edges, since no marked edge is in $D^p_{j,k,l}$.
But then $D^p_{j,k,l}$ must include one edge to cover the top vertex of $T^p_{j,k,l}$ and one edge to cover the bottom vertex of $T^p_{j,k,l}$, thus concluding the proof.
\end{proof}

\begin{theorem}
\label{NPtau}
\TEdgeCover{} for graphs of lifetime $2$ is NP-complete. 
\end{theorem}
\begin{proof}
First, \TEdgeCover{} is in NP, since
we given a set $E'$ of edges, $|E'| \leq k$, we can decide in polynomial time if $E'$ covers each temporal vertex
of the input temporal graph.

Given an instance $F$ of \threeSAT{(2,2)}, 
we have described how to construct a corresponding temporal graph $\mathcal{G}$ with lifetime $2$.
We claim that $F$ is satisfiable if and only if there exist a covering of the temporal vertices of $\mathcal{G}$ that uses at most $5n+6m$ edges.

$(\Rightarrow)$. Assume that $F$ is satisfiable, and let $\sigma$ be an assignment that satisfies $F$. Then we construct an edge cover $E'$ of $\mathcal{G}$
as follows.
For each variable $x_i$, $i \in [n]$,
if $\sigma(x_i)$ is true, we add $E'_{i,1}$ to $E'$ 
(five edges including the edges marked $i$), 
if $\sigma(x_i)$ is false, we add $E'_{i,2}$ to $E'$ 
(five edges including the edges marked $-i$). 
In this way we add $5n$ edges to $E'$.

Now, for each clause $C_p$, with $p \in [m]$, we know that, since $F$ is satisfied by $\sigma$, we have added to $E'$ at least one of the marked edges of $C^p_{j,k,l}$. Thus, by \cref{6coverLemma}, we can use six edges to cover the remaining temporal vertices of $T^p_{j,k,l}$. In this way we add $6m$ edges to $E'$.
Thus $E'$ contains $5n+6m$  edges and covers all the temporal vertices of $\mathcal{G}$.

$(\Leftarrow)$. Suppose that there exist an edge cover $E'$ of the temporal vertices of $\mathcal{G}$ of cardinality at most $5n+6m$. 
First, note that $E'$ must use no less than $5n+6m$ edges to cover the temporal vertices of $\mathcal{G}$. 
Indeed, the temporal vertices of each $L_i$, $i \in [n]$, needs at least five edges, since 
$L_i$ is a cycle with ten vertices.
Moreover, by \cref{6coverLemma} and \cref{6coverLemmab}, the vertices of each $T^p_{j,k,l}$, $p \in [m]$ and $j,k,l \in [n]$, need at least 6 edges (excluding the marked edges). 

Now, each $L_i$, $i \in [n]$, is covered by exactly five edges of $E'$, that is either $E'_{i,1} \subseteq E'$ or $E'_{i,2} \subseteq E'$.
Then each $T^p_{j,k,l}$, $p \in [m]$ and $j,k,l \in [n]$, is covered with exactly six edges of $E'$, excluding the marked edges.
Moreover, at least one of these marked edges must be in $E'$, as a consequence of \cref{6coverLemmab}.

Now, we define an assignment for $F$. 
For each variable $x_i$, $i \in [n]$, we define $\sigma(x_i)$ to be true if the edges of $L_i$ marked by $i$ are in $E'$, and false if the edges of $L_i$ marked by $-i$ are in $E'$ instead.
Then $\sigma$ satisfies $F$. Indeed, $\sigma$ is well defined because either both edges marked by $i$ or both edges marked by $-i$ are in $E'$. Also, for each clause with a corresponding graph $T^p_{j,k,l}$, $p \in [m]$ and $j,k,l \in [n]$, at least one of the marked edges is in $E'$, thus implying that $\sigma$ satisfies a literal in the clause. Then $F$ is satisfied.

The NP-hardness of \TEdgeCover{} follows from the NP-hardness of \threeSAT{(2,2)}~\cite{DBLP:journals/dam/DarmannD21}.
\end{proof}

\subsection{Hardness when the Underlying Graph is a Tree}
\label{subsec:EdgeCoverTreeHard}

We show that \TEdgeCover{} is NP-complete when the underlying graph is a tree by giving a reduction from \SetCover{} to \TEdgeCover{}.

\begin{figure}[t]
\centering
    \includegraphics[scale=0.8]{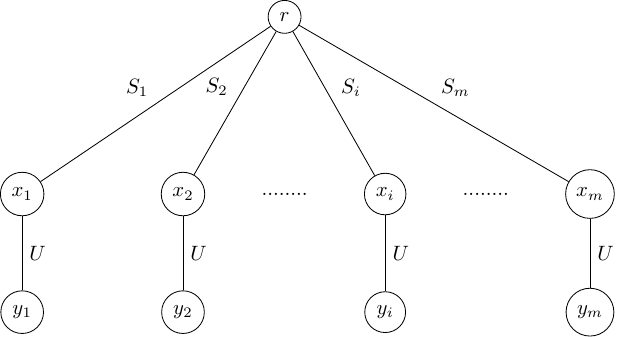}
\caption{The temporal graph obtained from an instance of \SetCover{}.}
\label{treeedgecover}
\end{figure}

Given an instance $(U,\mathcal{S},k)$ of \SetCover{}, where $U=[n]$ and $\mathcal{S}$ consists of $m$ sets $S_1$,\dots, $S_m$ ($S_i \subseteq [n]$, for each $i \in [m]$, ), we construct a corresponding temporal graph $\mathcal{G}$ (see \cref{treeedgecover}).
$\mathcal{G}$ has an underlying graph $G$ which is is a tree rooted in $r$; $r$ has $m$ children $x_1$, \dots, $x_m$, and each $x_i$ has a single child $y_i$, with $i \in [m]$.
Function $\lambda$ associates time label to each edge as follows: $\lambda(x_i y_i) = S_i$ and $\lambda(r x_i) = U$, for each $i \in [m]$.
 The idea of the reduction is that each edge $x_i y_i$, $i \in [m]$, must be in a temporal edge cover, and that the temporal vertices 
 $(r,j)$, $j \in [m]$, are covered by edges incident in $r$
that encode a set cover.

\begin{theorem}
\label{NPtree}
\TEdgeCover{} is NP-complete even when the underlying graph is a tree.
\end{theorem}
\begin{proof}
As discussed in the proof of \cref{NPtau},
\TEdgeCover{} is in NP.
We present a reduction from \SetCover{} to \TEdgeCover{} tree.
Given an instance $(U,\mathcal{S},k)$ of \SetCover{}, where $U=[n]$ and $\mathcal{S}$ consists of $m$ sets $S_1$,\dots, $S_m$ ($S_i \subseteq [n]$, for each $i \in [m]$, ), we construct a corresponding temporal graph $\mathcal{G}$ (see \cref{treeedgecover}).
$\mathcal{G}$ has an underlying graph $G$ which is is a tree rooted in $r$; $r$ has $m$ children $x_1$, \dots, $x_m$, and each $x_i$ has a single child $y_i$, with $i \in [m]$.
Function $\lambda$ associates time label to each edge as follows:

\begin{itemize}
    \item For each $i \in [m]$, $\lambda(x_i y_i) = S_i$
    \item For each $i \in [m]$, $\lambda(r x_i) = U$ 
\end{itemize}


We show next that there exists a covering of $U$
with at most $k$ sets if and only if 
$\mathcal{G}=(G,\lambda)$ has an edge cover of at most $k+m$ edges.

$(\Rightarrow)$. Let $S_{i_1}$, \dots, $S_{i_\ell}$ be sets in $\mathcal{S}$, 
such that $\bigcup_{j=1}^{\ell} S_{i_j} = U$, with $\ell\le k$. Then consider the following set $E'$ of $m+\ell$ edges: 
\[
E'= \bigcup_{i=1}^m \{x_i y_i\} \cup
\bigcup_{j=1}^{\ell} \{r x_{i_j}  \}
\] 

$E'$ covers all the temporal vertices of $\mathcal{G}$.
Indeed, edge $x_i y_i$ covers each $(x_i,t)$ and each $(y_i,t)$, $i \in [m]$ and $t \in [n]$. The temporal vertices $(r,t)$, $t \in [n]$, are covered by edges $r x_{i,j}$, since, by hypothesis, $\bigcup_{j=1}^{\ell} S_{i_j} = U$.
Since $m+\ell \le m+k$, we have obtained an edge cover of the desired size.

$(\Leftarrow)$. Consider an edge covering $E'$ of $\mathcal{G}$, with $|E'| = m+\ell$ edges, where $\ell \le k$. Note that $E'$ must contain each edge $x_i y_i$, $i \in [m]$, otherwise it would be impossible to cover the temporal vertices $(y_i,t)$,
$t \in [n]$. This implies that $E'$ contains $\ell$ edges $r x_{i_1}, \dots, r x_{i_\ell}$. Since these edges cover each temporal vertex $(r,t)$, with $t \in [n]$, then for the corresponding sets $S_{i_j}$, $j \in [\ell]$, it holds that $\bigcup_{j=1}^{\ell} S_{i_j} = U$, and we obtain a solution to \SetCover{} with at most $k$ sets.

The NP-hardness of \TEdgeCover{} when the underlying graph is a tree follows from the NP-hardness of \SetCover{}~\cite{DBLP:conf/coco/Karp72}.
\end{proof}

\subsection{\FPT{} algorithm in $\tau$ and treewidth for \TEdgeCover{}}
\label{FPTECsection}

In this subsection we present an \FPT{} algorithm that finds the minimum cardinality of a temporal edge cover of $\mathcal{G}$.
Note that we can assume, without loss of generality, that each temporal vertex of the temporal graph $\mathcal{G}$ can be covered by at least one edge. That is, we can assume that there are no independent temporal vertex in $\mathcal{G}$, since those would not need to be covered and can be ignored during the computation.

Let $\mathcal{G}=(G,\lambda)$ be a temporal graph and consider a nice tree decomposition $(T,\{X_t\}_{t\in V(T)})$, with $T$ rooted at $r$, of a $G$. For each $t\in V(T)$, let $G_t$ be the subgraph of $G$ containing all the vertices $v\in X_{t'}$ for any $t'$ in the subtree rooted at $t$. Also, for any $X\subseteq V(G)$, let $E(X)$ denote the set of edges with some endpoint in $X$ (formally, $E(X)=\{uv \in E(G) \mid u,v\in X\}$). 

Given $R\subseteq V(G)$, we denote by $V^T(R)$ the set of temporal vertices $R\times [\tau]$; for simplicity, we write $V^T(G)$ to denote $V^T(V(G))$.  
Given $S\subseteq E(G)$, we denote by $V^T(S)$ the set of temporal vertices which are endpoints of $S$, i.e., $V^T(S) = \bigcup_{e\in S}\{(u,i)\mid i\in \lambda(e), u\in e\}$. 
Additionally, given $S\subseteq E(G)$ and $(u,i)\in V^T(G)$, we say that \emph{$S$ covers $(u,i)$} if there exists $e\in S$ such that $u\in e$ and $i\in \lambda(e)$. Observe that $S$ covers $V^T(S)$.

As is usual the case when using tree decomposition, we work with partial solutions, i.e., with sets of edges that only partially cover the temporal vertices of $G_t$. This is because we might cover some temporal vertex $(u,i)\in X_t\times[\tau]$ only with an edge introduced later, i.e., with an edge $uv$ such that $v\notin V(G_t)$. Therefore, for each node of $T$, we keep track of the temporal vertices within $X_t\times[\tau]$ that are covered and of the edges within $E(X)$ that are chosen. Formally, 
given $t\in V(T)$, for each $S\subseteq E(X_t)$ and each $C\subseteq X_t\times [\tau]$ with $V^{T}(S)\subseteq C$, we define:

\[\begin{array}{rll}
T_t(S,C) = & 
              \min \{k\mid & \text{there exists $S'\subseteq E(G_t)$ with $\lvert S'\rvert=k$ s.t. $S'\cap E(X_t)=S$  }\\ 
           &               & \text{and the set of temporal vertices in $V^T(G_t)$}\\
           &               & \text{covered by $S'$ is exactly $C\cup V^T(G_t\setminus X_t)$}\}
\end{array} \]



If there is no such set $S'$, then $T_t(S,C)=+\infty$. 
Essentially, the function gives the minimum cardinality of a partial edge cover $S'$ for the temporal graph $(G_t, \lambda\restriction_{E(G_t)})$ such that:
\begin{itemize}
\item $S$ is exactly the set of edges in $E(X_t)$ that are selected by $S'$; 
\item $C$ is exactly the set of temporal vertices in $X_t\times [\tau]$ covered by $S'$. Observe that these must include the endpoints of the temporal edges related to the edges selected in $S$, and this is why we ask for $V^T(S)$ to be contained in $C$; and
\item Each temporal vertex related to some vertex in $G_t\setminus X_t$ must be covered by $S'$.
\end{itemize}

Observe that $T_r(\emptyset,\emptyset)$ gives us the minimum cardinality of a temporal edge cover for $\mathcal{G}$. In what follows, we show how to recursively compute $T_t(S,C)$ for each $t\in V(T)$, $S\subseteq E(X_t)$, and $C\subseteq (X_t\times [\tau))$ with $V^T(S)\subseteq C$, depending of the type of node $t$.

\begin{itemize}
\item leaf: if $t$ is a leaf, then $T_t(\emptyset,\emptyset)=0$;
\item introduce node: let $v\in V(G)$ be the vertex introduced by $t$ and $t'$ be its only child. Also, let $D$ be the set of temporal vertices $(u,i)$ with $u\neq v$ covered by some edge incident to $v$. Formally, 
$D=V^{T}(S\cap\delta_{G}(v))\setminus (\{v\}\times [\tau])$. 
Additionally, let $S'=S\setminus \delta_G(v)$, $C'= C\setminus (\{v\}\times [\tau])$, and $k=|S\cap\delta_{G}(v)|$. 
We have that:
\[
T_t(S,C) = \left\{
\begin{array}{ll}
     k+\min_{\hat{D}\subseteq D}T_{t'}(S', C'\setminus \hat{D})  &  \text{, if } V^{T}(S)\cap (\{v\}\times [\tau]) = C\cap (\{v\}\times [\tau]) \\
     +\infty & \text{, otherwise}
\end{array}
\right.\]


\item forget node: let $v\in V(G)$ be the vertex forgotten by $t$ and let $t'$ be its only child. 
Also, let $S' = \delta_G(v)\cap E(X_{t'})$. Then:
\[
T_t(S,C) = \min_{\hat{S}\subseteq S'}T_{t'}(S\cup\hat{S},C\cup (\{v\}\times [\tau])).
\]
%
%
\item join node: let $t_1$ and $t_2$ be the two children of $t$. By definition, we know that $X_{t_1}=X_{t_2}$. Then:
\[
T_t(S,C)=-|S| +\min\{T_{t_1}(S,C_1) + T_{t_2}(S,C_2)\mid C_1\cup C_2 = C\text{ and } V^{T}(S)\subseteq C_1\cap C_2\}.
\]
\end{itemize}




\begin{theorem}
\label{thm:FPTEdgeCover_twplustau}
\TEdgeCover{} can be computed in time $O^*(2^{w^2}\cdot 8^{w\cdot \tau})$.
\end{theorem}
\begin{proof}
Recall that $X_r=\emptyset$, where $r$ is the root of the tree decomposition. Hence the only entry related to $r$ is $T_r(\emptyset,\emptyset)$. By definition, this entry is equal to the minimum value $k$ for which there exists $S'\subseteq E(G_r) = E(G)$ such that $\lvert S'\rvert = k$ and, for every $(u,i)\in V(G_r)\times [\tau] = V(G)\times[\tau]$, there exists $e\in S'$ with $u\in e$ and $i\in \lambda(e)$. In other words, $T_r(\emptyset,\emptyset)$ is equal to the minimum size of an edge cover of $\mathcal{G}$. It therefore remains to show that each entry $T_t(S,C)$ is computed correctly and that it takes $O^*(2^{w^2}\cdot 8^{w\cdot \tau})$ time to compute the entire table $T$. We start by analyzing each possible type of a node $t$.

\begin{itemize}
    \item  $t$ is a leaf, then we can only have $S = \emptyset = C$, and there is only one subset of $E(G_t)$, which is again the empty set and has cardinality $0$. Therefore $T_t(\emptyset,\emptyset)=0$.

\item $t$ is an introduce node: let $v$ be the vertex introduced by $t$ and let $t'$ be its only child. 
Recall that, since $(T,\{X_t\}_{t\in V(T)})$ is a nice tree decomposition, each node of $G$ is forgot precisely once, and for each edge of $G$ there exists an note in the tree that contains both the vertices adjacent to that edge. Therefore all the edges in $E(G_t)$ that are adjacent to $v$ are also contained in $E(X_t)$. This means that $S$ must cover the temporal vertices $(v,i)$ contained in $C$. Therefore if $V^{T}(S)\cap (\{v\}\times [\tau]) \neq C\cap (\{v\}\times [\tau])$, then there cannot be any temporal edge cover that satisfies the properties required by $T_t(S,C)$. In such case we get $T_t(S,C)=+\infty$.

Now suppose that $V^{T}(S)\cap (\{v\}\times [\tau]) = C\cap (\{v\}\times [\tau])$. 
As before, let $D=V^{T}(S\cap\delta_{G}(v))\setminus (\{v\}\times [\tau])$, $S' = S\setminus\delta_G(v)$, and $C' = C\setminus (\{v\}\times [\tau])$. 
Also, let $\hat{D}\subseteq D$ be such that $T_{t'}(S', C'\setminus \hat{D})$ is minimum. 
We want to show that the associated temporal edge cover gives a minimum temporal edge cover associated to $T_t(S,C)$, and vice versa. Specifically, we show that we can obtain one from the other by adding or removing the set of edges $S\cap\delta_{G}(v)$.

By definition of $T_{t'}$, there exists a set $R'\subseteq E(G_{t'})$ such that $R'\cap E(X_{t'}) = S'$ and $R'$ covers exactly $C'\setminus \hat{D}$ and all the temporal vertices in $V(G_{t'}\setminus X_{t'})\times [\tau]$. Define $S''=R'\cup (S\cap\delta_{G}(v))$. 
First, note that $\lvert S''\rvert = k+T_{t'}(S',C'\setminus \hat{D})$, as indeed $R'$ cannot contain any edge incident to $v$ since $v\notin V(G_{t'})$. By construction we also get that $S''\cap E(X_t) = S$. It thus remains to argue that the set of temporal vertices in $V(G_t)\times [\tau]$ covered by $S''$ is exactly equal to $C\cup (V(G_t\setminus X_t)\times [\tau])$. 
Note that the only such temporal vertices not covered by $R'$ are exactly the copies of $v$ covered by $\delta(v)\cap S$. 

Therefore
\[
T_t(S,C) \le k + \min_{\hat{D}\subseteq D}\{T_{t'}(S', C'\setminus \hat{D}).
\]
To prove that the two sides are equal we just need to reason in the inverse direction. 
Given $S''\subseteq E(G_t)$ a subset that minimizes entry $T_t(S,C)$, we define $R'=S''\setminus (S\cap\delta_{G}(v))$ and show that $R'$ satisfies the conditions within the definition of $T_{t'}(S',C'\setminus \hat{D})$. 
Since the equation takes the minimum over $\hat{D}$, then 
\[
T_t(S,C) \ge k + \min_{\hat{D}\subseteq D}T_{t'}(S', C'\setminus \hat{D}).
\]
and the equality holds.

\item $t$ is a forget node: let $v$ be the vertex forgotten by $t$ and let $t'$ be its only children. We prove that
\[
T_t(S,C) = \min_{\hat{S}}T_{t'}(S\cup\hat{S},C\cup (\{v\}\times [\tau])),
\]
where $\hat{S}$ varies over all the sets of edges adjacent to $v$ in $X_t$, i.e. $\hat{S}\subseteq \delta_{X_{t'}}(v)$. 
By definition, $T_t(S,C)$ is the minimum cardinality of a set $S'\subseteq E(G_t)$ that covers exactly the temporal vertices $C \cup (V(G_t\setminus X_t)\times [\tau])$ and such that $S'\cap E(X_t)=S$. On the other hand, $T_{t'}(S\cup\hat{S},C\cup (\{v\}\times [\tau]))$, for some $\hat{S}\subseteq\delta_{X_{t'}}(v)$, is the minimum cardinality of a set $R'\subseteq E(G_{t'})$ such that $R'\cap E(X_{t'})=S\cup\hat{S}$ and that covers exactly the temporal vertices in $C\cup (\{v\}\times [\tau])$ and $V(G_{t'}\setminus X_{t'})\times [\tau]$. That is, $R'$ covers $C$, $\{v\}\times [\tau]$ and $V(G_{t'}\setminus X_{t'})\times [\tau]$. Since $G_t = G_{t'}\setminus \{v\}$, then $R'$ covers $C$ and $V(G_{t}\setminus X_{t})\times [\tau]$.  
Also, $R'\cap E(X_t) = R'\cap (E(X_{t'}\setminus \delta_{G}(v)) = (R'\cap E(X_{t'}))\setminus \delta_{G}(v) = (S\cup\hat{S})\setminus \delta_{G}(v) = S$.
Therefore
\[
T_t(S,C) \le \min_{\hat{S}}T_{t'}(S\cup\hat{S},C\cup (\{v\}\times [\tau])),
\]
Proving the opposite is easy, as any set $S''$ that satisfies the conditions of $T_t(S,C)$ also satisfies those of $T_{t'}(S\cup\hat{S},C\cup (\{v\}\times [\tau]))$, with $\hat{S}=S''\cap \delta_{G}(v) \cap E(X_{t'})$.

\item $t$ is a join node: let $t_1$ and $t_2$ be the children of $t$. 
We prove that
\[
T_t(S,C)=-|S| +\min\{T_{t_1}(S,C_1) + T_{t_2}(S,C_2)\mid C_1\cup C_2 = C\text{ and } V^{T}(S)\subseteq C_1\cap C_2\}.
\]

Let $C_1$, $C_2$ be any pair of sets such that $C_1\cup C_2 = C $ and $V^{T}(S)\subseteq C_1\cap C_2$. 
Suppose that $T_{t_1}(S,C_1) + T_{t_2}(S,C_2)$ is smaller than $\infty$. 
Then there exist two sets $S'_1\subseteq E(G_{t_1})$ and $S'_2\subseteq E(G_{t_2})$ such that, for each $i\in[2]$, we have that $S'_i\cap E(X_{t_i}) = S$ and $S'_i$ covers exactly $C_i\cup V^T(G_{t_i}\setminus X_{t_i})$. Observe that this directly gives us that $S'=S'_1\cup S'_2$ covers exactly $C\cup V^T(G_t\setminus X_t)$. 

Now, since in a tree decomposition each vertex is forgot exactly once and the nodes that contain a vertex form a connected component, then the only vertices in common between $G_{t_1}$ and $G_{t_2}$ are those in $X_t$. Therefore the only edges contained both in $S'_1$ and $S'_2$ are those of $S$. Thus $|S'|=|S'_1|+|S'_2|-|S|$, and
\[
T_t(S,C)\le\min_{C_1,C_2}\{T_{t_1}(S,C_1) + T_{t_2}(S,C_2)\}-|S|.
\]

On the other hand, any temporal edge $S'$ cover associated to $T_t(S,C)$ gives two temporal edge covers $S'_1=S'\cap E(G_{t_1}$ and $S'_2 = S'\cap E(G_{t_2})$. Specifically, there exists two sets $C_1,C_2\subseteq C$, with $C_1\cup C_2=C$, such that $S'_1$ covers all and only the temporal vertices $C_1\cup (G_t\setminus X_t \times [\tau])$ and $S'_2$ covers all and only $C_2\cup (G_t\setminus X_t \times [\tau])$. Hence
\[
T_t(S,C)\ge\min_{C_1,C_2}\{T_{t_1}(S,C_1) + T_{t_2}(S,C_2)\}-|S|,
\]
and the equality holds.

\end{itemize}

Now we analyse the running time needed to compute the given recursive function. It is known that a tree decomposition with $O(n)$ nodes can be assumed, where $n=\lvert V(G)\rvert$ (see e.g.~\cite{cygan2015parameterized}). Hence, we just need to compute, for each node $t$, the size of $T_t$ and the running time needed to compute an entry of $T_t$. 
So consider $t\in V(T)$. First note that there are at most $2^{\binom{|X_t|}{2}}$ subsets $S\subseteq E(X_t)$ and $2^{|X_t\times [\tau]|}$ subsets $C\subseteq X_t\times [\tau]$. Therefore $T_t$ has $O(2^{w^2}\cdot 2^{w\cdot\tau})$ entries. Now, we analyze the the running time needed to compute an entry of $T_t$, depending on the type of node $t$. 

\begin{itemize}
    \item $t$ is a leaf: then the smallest $T_t(\emptyset,\emptyset)=0$, and this is computed in $O(1)$.

\item $t$ is an introduce node: then checking whether $V^{T}(S)\cap (\{v\}\times [\tau]) = C\cap (\{v\}\times [\tau])$ takes time $O(\tau)$. Also there are $O(2^{w})$ possible subsets $\hat{D}\subseteq D$ for which we need to check the values of $T_{t'}$. Since we assume to have already computed $T_{t'}$, the computation of $T_{t}(S,C)$ for an introduce node takes time $O^*(2^{w})$ ($\tau$ remains hidden in the $O^*$ notation).

\item $t$ is a forget node: applying similar reasoning, observe that it takes time $O(2^{w})$ (which is the number of subsets $\hat{S}\subseteq \delta_{X_{t'}}(v)$).

\item $t$ is a join node: similarly, we count the number of combinations for $C_1$ and $C_2$, which gives $O(2^{2 w \cdot \tau}) = O(4^{w\cdot \tau})$.
\end{itemize}

The worst case is the one for the join node, which takes $O(4^{w \cdot \tau})$ time. Multiplying by the size of $T_t$, the theorem follows. 
\end{proof}

\section{Approximation of \TEdgeCover{}}
\label{sec:ApproxEdge}

In this section we consider the approximability
of \TEdgeCover{}. First, we show a bound on the approximation ($b \log \tau$, for any constant 
$0 <b <1$), then we present an approximation algorithm of factor $O(\log \tau)$.

\subsection{Inapproximability} 

\begin{figure}[t]
\centering
\includegraphics[scale=0.9]{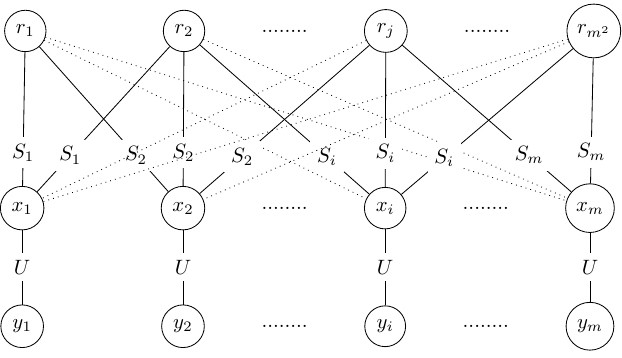}
\caption{The temporal graph obtained from an instance of \SetCover{}. Some edges are dotted for readability, and we are not showing the labels on those edges for the same reason.}
\label{fig:inapproximability}
\end{figure}

We show that \TEdgeCover{} cannot be approximated
within factor $b \log \tau$, for any constant $0 < b < 1 $.
We prove this result by giving an approximation preserving reduction from the \SetCover{} problem\footnote{In this section we consider the
optimization version of \SetCover{}, thus we omit $k$ from the instance of the problem.}.
Consider an instance $(U,\mathcal{S})$ of \SetCover{}, where
$U = \{ u_1, \dots , u_n\}$  and  
$\mathcal{S} = \{ S_1, \dots , S_m\}$.
We can assume $U=[n]$, therefore each $S_i$, $i \in [m]$ is a subset of $[n]$.
We define a corresponding instance $\mathcal{G}=(G, \lambda)$ of \TEdgeCover{} as follows (see \cref{fig:inapproximability}):
\begin{align*}
V(G)&= \{ r_i\mid i \in [m^2]\} \cup \{ x_i\mid i \in [m] \}
\cup \{ y_i\mid i \in [m] \}\\
E(G)&= \{r_i x_j \mid i \in [m^2], j \in [m]\} \cup \{x_i y_i \mid i \in [m] \}\\
\lambda&: E(G)\rightarrow \mathcal{P}([n]), \quad
\lambda(e)=
\begin{cases}
S_j \qquad \text{if $e=r_i x_j$ for some $i\in [m^2]$, $j\in m$,}\\
U \qquad \text{if $e=x_i y_i$ for some $i \in [m]$.}\\
\end{cases}
\end{align*}

Note that $\mathcal{G}$ has lifetime $\tau=n$.
We now show the main properties of the reduction.

\begin{lemma}
\label{lem:inapprox1}
Consider an instance $(U,\mathcal{S})$ of \SetCover{}
and a corresponding instance $\mathcal{G}$ of \TEdgeCover{}.
Given a solution $\mathcal{S}'$ of \SetCover{} on instance $(U,\mathcal{S})$, we can
compute in polynomial time a solution of \TEdgeCover{} on instance
$\mathcal{G}$ that consists of at most $m + |\mathcal{S}'| m^2$ edges.
\end{lemma}
\begin{proof}
Given solution $\mathcal{S}$ of \SetCover{} on instance $(U,\mathcal{S})$, we define an edge cover $E'$ of $\mathcal{G}$ as follows:

\begin{itemize}
    \item for each $i \in [m]$, each edge $x_i y_i$ is added to $E'$, thus temporal vertices $x_i t$ and $y_i t$ are covered, for each $t \in \tau$;
    \item for each $j \in [m^2]$, add edge $r_j x_i$, with $S_i \in \mathcal{S}'$, to $E'$; since the sets in $\mathcal{S}'$ cover $U$, it follows that for each $u_t \in U$ there exists a set $S_h \in \mathcal{S}'$ such that $u_i\in S_h$. Thus each temporal vertex $(r_j, t)$ is covered by the edge $r_j x_h\in E'$.
\end{itemize}
The number of edges in $E'$ is $m + |\mathcal{S}'| m^2$ edges, thus the lemma follows.
\end{proof}

\begin{lemma}
\label{lem:inapprox2}
Consider an instance $(U,\mathcal{S})$ of \SetCover{} and a corresponding instance $\mathcal{G}$ of \TEdgeCover{}.
Given a solution $E'$ of \TEdgeCover{} on instance $\mathcal{G}$, then there exists a positive integer $k$ such that $|E'|= m + k m^2$.
Then we can compute in polynomial time a solution of \SetCover{} on instance $(U,\mathcal{S})$ that consists of at most $k$ sets.
\end{lemma}
\begin{proof}
Given solution $E'$ of \TEdgeCover{}, we start by proving some properties of $E'$.
First, consider vertices $y_i$, with $i \in [m]$.
Since each $y_i$ has degree one, then each edge $x_i y_i$ must be in $E'$, and it covers all the temporal vertices $(y_i, t)$ and $(x_i, t)$.
Define $E'' \subseteq E'$ as $E'' = \bigcup_{i \in [m]} x_i y_i$.

Now, consider the vertices $r_j$, with $j \in [m^2]$, and consider the edges in $E'$. 
Note that $E' = E'' \cup \bigcup_{j \in [m^2]} E'_j$, where, for each $j \in [m^2]$, we define $E'_j = \{r_j x_i\in E'\mid i\in [m]\}$.
Note that the sets $E'_j$ are disjoint.
We show that they have the same cardinality.
Let $E'_h$ be a set of minimum cardinality among the sets $E'_j$, $j \in [m^2]$, and define the sets $\hat{E}'_j = \{ r_j x_i \mid r_h x_i \in E'_h\}$. Since $E'_h$ covers all the temporal vertices in $\{r_h\}\times [n]$, then $\hat{E}'_j$ covers the temporal vertices of $\{r_j\}\times [n]$ for each $j=1,\dots,m$. 
Thus $E'' \cup \bigcup_{j \in [m^2]} \hat{E}'_j$ is a solution of \TEdgeCover{} and has cardinality $m + |E'_h| m^2$. Since $E'$ is a solution and $E'_h$ has minimum cardinality among the $E'_j$'s, then $|E'| = m + |E'_h| m^2$ and all the $E'_j$ have the same cardinality. This proves the first statement of the lemma, with $k=|E'_h|$.

Now define a set cover $S'$ as follows:

\begin{itemize}
    \item for each edge $r_h x_i \in E'_h$, with $i \in [m]$, add set $S_i$ to $S'$
\end{itemize}

Since $E'_h$ contains $k$ edges, then $S'$ contains $k$ sets.

Now, we prove that $S'$ covers each element in $U$.
Consider an element $u_j$, $j \in [n]$.
Since $E'$ is an edge cover, it covers each temporal vertex $(r_h, j)$, thus there exists an edge $r_h x_i$ in $E'$, which hence belongs in $E'_h$, such that $j\in\lambda(r_h x_i)$.
By construction, $S'$ contains set $S_i$, such that $u_j \in S_i$, thus concluding the proof.
\end{proof}

\begin{theorem}
\TEdgeCover{} cannot be approximated within factor $b \log \tau$, for any $b$ with $0 <b<1$, unless $P = NP$.
\end{theorem}
\begin{proof}
Let $(U,\mathcal{S})$ be an instance of \SetCover{}, and $\mathcal{G}$ the corresponding instance of \TEdgeCover{}.
Consider an approximated (optimal, respectively) solution $E'_A$ ($E'_O$, respectively) of \TEdgeCover{} on instance $\mathcal{G}$.
Consider the approximation factor of \TEdgeCover{}: $\frac{|E'_A|}{|E'_O|}$.

By Lemma \ref{lem:inapprox2}, we can compute in polynomial time an approximated solution $\mathcal{S}'_A$ of \SetCover{} on instance $(U,\mathcal{S})$, such that
\[
\frac{|E'_A|}{|E'_O|} \geq \frac{|\mathcal{S}'_A|m^2+m}{|E'_O|}.
\]
By Lemma \ref{lem:inapprox1}, for
an optimal solution $\mathcal{S}'_O$ of \SetCover{} on instance $(U,\mathcal{S})$, it holds that $|E'_O| \leq |\mathcal{S}'_O|m^2 + m.$

By combining the two inequalities, it holds that
\[
\frac{|E'_A|}{|E'_O|} \geq \frac{|\mathcal{S}'_A|m^2+m}{|E'_O|} \geq
\frac{|\mathcal{S}'_A|m^2+m}{|\mathcal{S}'_O|m^2+m}.
\]
Thus 
\[
\frac{|E'_A|}{|E'_O|} \geq
\frac{|S'_A|m^2+m}{|\mathcal{S}'_O|m^2+m} \geq \frac{|\mathcal{S}'_A|m^2}{|\mathcal{S}'_O|m^2+m} \geq \frac{|\mathcal{S}'_A|m^2}{|\mathcal{S}'_O|m^2} \cdot \frac{|\mathcal{S}'_O|m^2}{|\mathcal{S}'_O|m^2+m}.
\]
It holds that
\[
\frac{|\mathcal{S}'_O|m^2}{|\mathcal{S}'_O|m^2+m} \geq 1 -o(1).
\]
Since \SetCover{} is not approximable within factor $c \ln |U|$, for any constant $c$ such that $0 < c <1$, unless $P = NP$~\cite{DBLP:journals/talg/AlonMS06,DBLP:conf/stoc/DinurS14}, then for any constant $b$, with $0 <b<1$, unless $P = NP$, it follows that
\[
\frac{|E'_A|}{|E'_O|} > b \ln |U| \qquad \text{for any constant $0<b<1$}.
\]
Since $|\tau| = |U|$, the theorem follows.
\end{proof}

\subsection{A $O(\log \tau)$-Approximation Algorithm}

In this section we present an approximation algorithm for \TEdgeCover{} of factor $O(\log \tau)$.
Given a temporal graph $\mathcal{G}=(G,\lambda)$, with lifetime $\tau$ and $G=(V,E)$, 
the algorithm assumes that the vertices are ordered -- the specific order is not relevant -- so we denote them as $v_1, \dots, v_n$.
The approximation algorithm, described in Algorithm \ref{Algo:ApproxLog}, computes an edge cover $E'$
by greedily covering the uncovered temporal vertices of each vertex $v_i$, $i \in [n]$, following the order (first it covers the uncovered temporal vertices of $v_1$, then of $v_2$ and so on, until all the temporal vertices are covered).
In order to cover the temporal vertices of each $v_i$, it applies the greedy algorithm of \SetCover{} on an instance that contains 
an element for each uncovered temporal vertex $(v_i,t)$ and a set, for each edge $v_i v_j\in E$,  that covers $(v_i, t)$ for each $t\in\lambda(v_i v_j)$.

More precisely, consider the $i$-th iteration,
$i\in [n]$, 
of Algorithm \ref{Algo:ApproxLog}.
Given the set $E'$ of edges computed by the first $i-1$-iterations of the algorithm, 
we define an instance $(U^i, \mathcal{S}^i)$ of \SetCover{}, where $U^i$ is the universe set and $\mathcal{S}^i$ is a collection of sets over $U^i$.
For each $i \in [n]$, the universe set $U^i$ is defined as
\[
U^i = \{t\in [\tau] \mid \text{$(v_i,t)$ is not covered 
by $E'$ and there exists a $v_i v_j\in E$ such that $t\in\lambda(v_i v_j)$}\}.
\]

The collection of sets $\mathcal{S}^i$ is defined as $\mathcal{S}^i = \{S^i_{e_1}, \dots, S^i_{e_z}\}$, where $e_1$,\dots, $e_z$ are the edges incident in $v_i$
and each $S^i_h \subseteq [\tau]$ is defined as $S^i_h = \{t\in [\tau] \mid t\in\lambda(e_h)\}$.

Algorithm \ref{Algo:ApproxLog} marks each temporal vertex as \emph{covered} when it adds to solution $E'$ an edge that covers it.

\begin{algorithm}[t]
\caption{Approximation algorithm for \TEdgeCover{} \\Input: a temporal graph $\mathcal{G}=(G,\lambda)$ with lifetime $\tau$.\\
Output: an edge cover $E'$ of $\mathcal{G}$ of approximation factor $O(\log \tau)$}
\label{Algo:ApproxLog}

Mark each temporal vertex $(v_i,t)$, $i \in [n], t \in [\tau]$ as \emph{uncovered}

$i\gets 1$\;

$E' \gets \emptyset$\;

    \ForEach{$i\in [n]$}{

        Define an instance $(U^i, \mathcal{S}^i)$ of \SetCover{} corresponding to $v_i$\;
        
        Compute (via a greedy approximation algorithm) an approximated solution $\mathcal{C}^i$ of \SetCover{} on instance $(U^i, \mathcal{S}^i)$\;
    
        Compute an approximation edge cover $E'_i$, by adding an edge $e_h$ to $E'_i$ if and only if $S^i_h \in \mathcal{C}^i$\;

        $E' \gets E' \cup E'_i$\;

        Mark each temporal vertex covered by $E'_i$ as \emph{covered}\;

        $i\gets i+1$\;
    }
Output $E'$ 
\end{algorithm}

Now, we show the correctness of Algorithm \ref{Algo:ApproxLog}.

\begin{lemma}
\label{lem:Algoapprox}
Let $E'$ be a solution computed by Algorithm \ref{Algo:ApproxLog}.
Then, denoted by $E^*$ an optimal solution of \TEdgeCover{} on instance $\mathcal{G}$, it holds that
\begin{enumerate}
    \item $E'$ is an edge cover of $\mathcal{G}$
    \item $|E'| \leq \log \tau \ |E^*|$.
\end{enumerate}
\end{lemma}
\begin{proof}
We start by proving that $E'$ is a feasible solution, then we prove that $|E'| \leq \log \tau |E^*|$.

1. By construction, at each iteration $i$, with $i \in [n]$, Algorithm \ref{Algo:ApproxLog} covers each temporal vertex associated with $v_i$ and not yet covered.
Indeed $E'_i$ contains an edge for each set in $\mathcal{C}^i$ and $\mathcal{C}^i$ covers $U^i$, which contains an element for each temporal vertex $(v,t)$ not covered at iteration $i$.
Hence each temporal vertex is eventually marked as \emph{covered} by Algorithm \ref{Algo:ApproxLog} and $E'$ covers each temporal vertex.

2. Consider solution $E'$ computed by Algorithm \ref{Algo:ApproxLog}, where $E' = \bigcup_{i=1}^n E'_i$ and $E'_i$, $i \in [n]$, is the set of edge computed by the $i$-th iteration.
Recall that $E'_i$ is computed from an approximated solution $\mathcal{C}^i$ of \SetCover{} on instance $(U^i,\mathcal{S}^i)$. By construction, $|E'_i|=|\mathcal{C}^i|$.

Denote by $OPT_i$ an optimal solution of \SetCover{} on instance $(U^i, \mathcal{S}^i)$.
Since $\mathcal{C}^i$ is computed by applying a greedy approximation of \SetCover{} on instance $(U^i, \mathcal{S}^i)$, and $|U^i| \leq \tau$, then $|E'_i| = |\mathcal{C}^i| \leq \log \tau \ |OPT_i|$.

Now, consider an optimal solution $H^* \subseteq E$ of \TEdgeCover{} on instance $\mathcal{G}$. For each $i \in [n]$, define the collection of sets $\mathcal{H}^i$ associated with $H^*$, $\mathcal{H}^i = \{S^i_{e_h}\in\mathcal{S}^i \mid e_h\in H^*\}$.

Note that $|H^*| = \sum_{i=1}^n \frac{1}{2}|\mathcal{H}^i|$ since, by construction, each edge $e_h \in H^*$
is incident in two vertices and thus 
$e_h = v_i v_j$ belongs to two collections of sets, 
namely $\mathcal{H}^i$ and $\mathcal{H}^j$.

The collection of sets in $\mathcal{H}^i \subseteq \mathcal{S}^i$ covers each element in $U^i$.
Indeed, $H^*$ covers each temporal vertex of $V$, hence also each temporal vertex incident in $v_i$, thus $\mathcal{H}^i$ contains at least one set for each $t \in U^i$.
Hence $\mathcal{H}^i$ is a feasible solution of \SetCover{} on instance $(U^i, \mathcal{S}^i)$ and, since $OPT_i$ is an optimal solution of \SetCover{} on instance $(U^i, \mathcal{S}^i)$, then $|OPT_i| \leq |\mathcal{H}^i|$.

Combining this with the formula for $|H^*|$, and $|E'_i| = |\mathcal{C}^i| \leq \log \tau  |OPT_i|$, we have
\[
|E'| = \sum_{i=1}^n |\mathcal{C}^i| \leq \log \tau  \ \sum_{i=1}^n |OPT_i| 
\leq  \log \tau \  \sum_{i=1}^n |\mathcal{H}^i|  = 2 \log \tau \ |H^*|,  
\]
thus concluding the proof.
\end{proof}

\section{\TMatching{}: Hardness and Tractability}
\label{sectionM}

In this section we consider the \TMatching{} problem and provide hardness results and tractability. The outline is the same as \TEdgeCover{}; we show that \TMatching{} is NP-complete when the lifetime $\tau$ of graph is 2, and then we show that it is NP-complete even when the underlying graph is a tree.
Finally, we describe an \FPT{} algorithm in $\tau$ and $w$ (treewidth) which solves \TMatching{}.

\subsection{Hardness for $\tau=2$}

\begin{figure}[t]
\begin{minipage}{0.20\linewidth}
    \centering
    \includegraphics[scale=0.9]{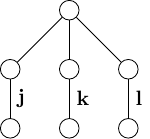}
    \caption{Graph $C_{j,k,l}$.}
    \label{Cjkl}
\end{minipage}
\hfill
\begin{minipage}{0.75\linewidth}
    \centering
    \includegraphics[scale=0.9]{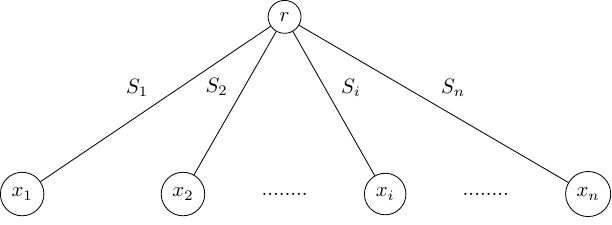}
    \caption{Temporal graph associated to an instance of \PSets.}
    \label{TreeMatchingimage}
\end{minipage}
\end{figure}

We show the NP-hardness of \TMatching{} restricted to lifetime 2 with reduction from $\threeSAT{(2,2)}$ similar to the one given in Section \ref{subsec:hard2}. This reduction follows the same idea as that of \cref{NPtau}. Indeed, we still use the graph $L_{i}$ defined in \cref{Lidef} and showed in \cref{Liimage}. 
For this reduction we do not encode clauses with
graphs isomorphic to $T_{j,k,l}$, but graphs
isomorphic to $C_{j,k,l}$, shown in \cref{Cjkl}. $C_{j,k,l}$ has three edges marked with integers $j$, $k$, $l$.


Let $F$ be an instance of $\threeSAT{(2,2)}$, with $n$ variables $x_1, \dots, x_n$ and $m$ clauses $C_1, \dots, C_m$.
We construct an associated temporal graph with lifetime $\tau=2$ defined in the following way. 
At time $1$, $\mathcal{G}$ contains a graph $G_1$ consisting
of the disjoint union of cycles $L_{i}$, $i \in [n]$, 
one for each variable $x_i$.
At time $2$, $\mathcal{G}$ contains a graph $G_2$ that
for each clause $C_p$ over variables $x_j$, $x_k$ and $x_l$, $j,k,l \in [n]$,
contains graph $C^p_{j,k,l}$ isomorphic to $C_{j,k,l}$. As in Section~\ref{subsec:hard2}, the marked edges
of $C^p_{j,k,l}$ are shared with cycles $L_i$, $i  \in \{j,k,l\}$ that encode 
the variables $x_j$, $x_k$ and $x_l$. The shared marked edge between $C^p_{j,k,l}$
and $L_i$ has mark $-i$ if $x_i$ is negated in the clause,
$i$ if the variable is positive in the clause. 
Note that $C^p_{j,k,l}$'s are build so that the marked edges of $L_i$
are in one-to-one correspondence with marked edges of $C_{j,k,l}$'s.

The correctness of the reduction follows from the fact that a maximum temporal matching of each $L_i$, $i \in [n]$, contains five edges, one including
positively marked edges and one including
negatively marked edges. This encodes an assignment to the
variables.
The temporal matching of each $C^p_{j,k,l}$ contains at most one unmarked edge.
However, a temporal matching $M$ of $\mathcal{G}$ 
contains one unmarked edge of $C^p_{j,k,l}$ only
if there is a marked edge shared by $C^p_{j,k,l}$ and
some $L_i$ that does not belong to $M$.
This encodes the fact that at least one literal of each clause must be 
satisfied. This reduction allows us to prove the following result.

\begin{theorem}
\label{teo:Match22hard}
\TMatching{} for graphs with lifetime 2 is NP-complete.    
\end{theorem}
\begin{proof}
First, \TMatching{} is in NP, since
we given a set $E'$ of edges, $|E'| \geq k$, we can decide in polynomial time if $E'$ is a temporal matching of the input temporal graph.

We now show the correctness of the reduction from \threeSAT{(2,2)} to \TEdgeCover{} previously described in this section.

Given an instance $F$ of \threeSAT{(2,2)}, we have a corresponding temporal graph $\mathcal{G}$ with lifetime $2$.
We claim that $F$ is satisfiable if and only if there exist a temporal matching of $\mathcal{G}$ that has at least $5n+m$ edges.

$(\Rightarrow)$. Assume that $F$ is satisfiable, and let $\sigma$ be an assignment that satisfies $F$, we construct a temporal matching $M$ as follows. For each variable $x_i$ , $i \in [n]$, $M$ contains five edges of $\mathcal{G}$ that are a temporal matching for $L_i$ as follows: if $\sigma(x_i)$ is false, $M$ contains a temporal matching of $L_i$ that includes the edges marked $i$, while if $\sigma(x_i)$ is true $M$ contains a temporal matching of $L_i$ that includes the edges marked $-i$. In this way $M$ contains $5n$ edges.

Now, for each clause $C_q$ with variables
$x_j$, $x_k$, $x_l$, we have that, since $F$ is satisfied by $\sigma$, for 
each subgraph $C^p_{j,k,l}$ at most two edges marked by $j$, $k$ and $l$ were added to $M$. Therefore we add to $M$ one of edges of $C^p_{j,k,l}$ incident in the top vertex. In this way we add $m$ edges, for a total of $5n+m$.

$(\Leftarrow)$. Consider a temporal matching $M\subseteq E(G)$ of $\mathcal{G}$ that contains at least $5n+m$ edges.
First, we prove some properties of $M$.
First note that, for $i \in [n]$, $M$ contains at most $5$ edges of $L_i$, otherwise the edges would share a vertex at time $1$. Since $M$ has cardinality at least $5n+m$, then at least $m$ edges do not belong to some $L_i$, 
$i \in [n]$. These $m$ edges must be edges of some subgraph $C^p_{j,k.l}$, $j,k,l \in [n]$, and must be unmarked,
since any marked edge belongs also to some $L_i$, $i \in [n]$.
Also note that for each $C^p_{j,k,l}$, at most one unmarked edge may belong to $M$, since otherwise the edges would share the top vertex at time $2$. Since there are $m$ clauses (hence $m$ subgraphs $C^p_{j,k,l}$), then exactly one unmarked edge of each $C^p_{j,k,l}$ belongs to $M$. 
This implies the following properties:

\begin{enumerate}
    \item For each cycle $L_i$, $i \in [n]$, either both 
edges marked $i$ or both edges marked $-i$ belong to $M$, otherwise
the edges in $M$ that belong to some $L_i$, $i \in [n]$, 
would be less than $5n$.

\item For each $C^p_{j,k,l}$ there exists at least one $q\in\{j,k,l\}$ such that both edges marked $q$ do not belong to $M$, otherwise no edge
of $C^p_{j,k,l}$ can be in $M$.

\end{enumerate}


We construct an assignment $\sigma$ as follows: for each $i \in [n]$, $\sigma(x_i)$ is false if $M$ does not contain the two edges marked $i$, and it is true if $M$ does not contain the two edges marked $-i$.
Then each clause is satisfied, because at least one the unmarked edge belongs to $M$, so at least one of the marked edges is not in $M$, so the corresponding literal is true.
\end{proof}

\subsection{Hardness when  the Underlying Graph is a Tree}
\label{sec:HardMatchTree}

We show a reduction from \PSets{} to \TMatching{}. 
Given an instance $(U,\mathcal{S}, k)$ of \PSets{} with $\mathcal{S}=\{S_1,\dots,S_n\}$ a collection of sets over a universe set $U$, we construct a temporal graph $\mathcal{G}=(G,\lambda)$ such that there exists $k$ disjoint sets in $\mathcal{S}$ if and only if there exists a temporal matching $M$ of $\mathcal{G}$ of size at least $k$.
Without loss of generality, we assume that $U =[n]$ and that each $S_i \subseteq [n]$, for each $i \in [m]$.

$\mathcal{G}=(G,\lambda)$ is defined as follows (the resulting temporal is presented in \cref{TreeMatchingimage}):
\begin{itemize}
\item $G$ is a tree rooted at a vertex $r$, which has $n$ children $x_1$, \dots, $x_n$
\item For each $ i\in [n]$, $\lambda(r x_i) = S_i$
\end{itemize}
The idea of the reduction is that since any pair of edges $r x_i$ and $r x_j$ of $\mathcal{G}$, where $i, j \in [n]$ and $i \neq j$,
share vertex $r$, then they can be in a temporal matching only if they 
are defined in different times, thus they are related to two disjoint
subsets $S_i$ and $S_j$ in an instance of \PSets{}.
Then, since \PSets{} is NP-complete~\cite{DBLP:conf/coco/Karp72}, we can prove the following result.


\begin{theorem}
\label{matchingtreeNP}
\TMatching{} is NP-complete even when the underlying graph $G$ is a tree.
\end{theorem}
\begin{proof}
As discussed in \cref{teo:Match22hard},
\TMatching{} is in NP.

We use now the polynomial reduction from \PSets{} to \TMatching{} described in this subsection to prove the NP-hardness of \TMatching{}.
We just need to prove that, given a collection $\mathcal{S}=\{S_1,\dots,S_n\}$ of sets and a nonnegative integer $k$, then there exists $l \geq k$ disjoint sets in $\mathcal{S}$ if and only if there exists a temporal matching $M$ of $\mathcal{G}$ of size at least $k$.

$(\Rightarrow)$: let $S_{i_1}$, \dots, $S_{i_l}$ be a disjoint collection of sets, with $l \geq k$
and $i_j \in [n]$
for each $j \in [l]$. Then we define $M=\{r x_{i_j} \mid j\in [l]\}$. That is, we take all the edges of $\mathcal{G}$ which are labeled with the sets of the disjoint collection. Clearly $|M|=l \geq k$, so we just need to prove that they are indeed a matching. That is, that the do not share any temporal vertex.

By contradiction, assume that there exists a time $t$ and two  $j_1, j_2 \in [n]$ such that $r x_{i_{j_1}}, r x_{i_{j_2}} \in M$ and $\{t\} \subseteq \lambda(r x_{i_{j_1}}) \cap \lambda(r x_{i_{j_2}})$. Then, by definition of the reduction, $t\in \lambda(r x_{i_{j_1}}) = S_{i_{j_1}}$ and $t\in \lambda(r x_{i_{j_2}}) = S_{i_{j_2}}$. Thus $t\in S_{i_{j_1}} \cap S_{i_{j_2}}$, which contradicts the fact that the $S_{i_j}$'s are disjoint. Therefore the elements of $M$ do not share any temporal vertex and $M$ is a matching, as required.

$(\Leftarrow)$: Let $M$ be a matching of $\mathcal{G}$ of cardinality $l \geq k$, then we can write $M$ as 
$\{r x_{i_1},\cdots, r x_{i_l}\}$, for some distinct integers $i_1$, \dots, $i_l \in [n]$. Then the corresponding collection $\{S_{i_1},\dots,S_{i_l}\}$ consists of $l \geq k$ distinct disjoint sets, 
as any two edges $r x_{i_{j_1}}$, $r x_{i_{j_2}}$
are defined in disjoint times.
This concludes the proof.
\end{proof}

\subsection{\FPT{} algorithm in $\tau$ and treewidth for \TMatching{}}

In this subsection we show an algorithm that finds the maximum cardinality of a temporal matching of $\mathcal{G}$ in \FPT{} time when parameterized by $\tau$ plus the treewidth. The approach follows the same idea as the \TEdgeCover{} one (see \cref{FPTECsection}).

Again, let $\mathcal{G}=(G,\lambda)$ be a temporal graph and consider a nice tree decomposition $(T,\{X_t\}_{t\in V(T)})$ of $G$, with $T$ rooted at $r$. We use the same notation as the one used in \cref{FPTECsection}. 
Given a matching $M\subseteq E(G)$ and a temporal vertex $(u,i)\in V^T(G)$, observe that $(u,i)$ can be covered by $M$ at most once, i.e., there is at most one edge $e\in M$ such that $u\in e$ and $i\in \lambda$. If such an edge exists, we say that $M$ \emph{saturates} $(u,i)$. 

We define the dynamic programming table $T_t$ related to each $t\in V(T)$ as follows. For each $N\subseteq E(X_t)$ and $C\subseteq V^T(X_t)$ with $V^{T}(N)\subseteq C$:

\[\begin{array}{ll}
     T_t(N,C)=  \max\{k \mid  & \exists \text{ a temporal matching } M\subseteq E(G_t) \text{ s.t. } |M|=k, \\
     &  M\cap E(X_t)=N\text{, and }V^{T}(M)\cap V^T(X_t) = C \}.\\
\end{array}
\]


If there exists no such set $M$ (e.g., it could happen that no temporal matching saturates exactly $C$), then $T_t(N,C)=0$. 
Essentially, the function gives the maximum cardinality of a temporal matching $M$ for the temporal graph $(G_t, \lambda\restriction_{E(G_t)})$ such that:
\begin{itemize}
\item $N$ is exactly the set of selected edges in $E(X_t)$;
\item $C$ is exactly the set of temporal vertices in $V^T(X_t)$ saturated by $M$.
\end{itemize}

Because $G_r = G$, the value of $T_r(\emptyset,\emptyset)$ tells us the maximum cardinality of a temporal matching for $\mathcal{G}$. We show how to recursively compute $T_t(N,C)$ for each $t\in V(T)$, $N\subseteq E(X_t)$, and $C\subseteq V^T(X_t)$, depending on the type of node $t$.

\begin{itemize}
\item leaf: if $t$ is a leaf, then $T_t(\emptyset,\emptyset)=0$;
\item introduce node: let $v\in V(G)$ be introduced by $t$ and let $t'$ be its only child. Also, let $F=N\cap\delta_{G}(v)$ be the set of edges in $N$ incident to $v$. 
Then: 
\[
T_t(N,C) = \left\{
\begin{array}{ll}
     T_{t'}(N\setminus F, C\setminus V^T(F)) + |F| & \text{, if }  V^T(F)\cap (\{v\}\times [\tau]) = C\cap (\{v\}\times [\tau])\\
     0 & \text{, otherwise.}
\end{array}
\right.\]

\item forget node: let $v\in V(G)$ be forgotten by $t$ and let $t'$ be its only child. 
To define the recursive function, let $\mathcal{N}$ contain every $\hat{N}\subseteq\delta_G(v)\cap E(X_{t'})$ such that $V^T(\hat{N})\setminus (\{v\}\times [\tau])\subseteq C$ and such that $\hat{N}$ is a matching. In words, it contains every subset of edges of $E(X_{t'})$ incident to $v$ whose other endpoints are temporal vertices within $C$, while also not having any edges sharing the same temporal vertices. Also, for any $\hat{N}\in\mathcal{N}$, let $\mathcal{C}_{\hat{N}}$ contain every $\hat{C}\subseteq \{v\}\times [\tau]$ such that $V^T(\hat{N}) \cap (\{v\}\times [\tau])\subseteq \hat{C}$. 
Then
\[
T_t(N,C) = \max\{T_{t'}(N\cup \hat{N},C\cup \hat{C})\mid  \hat{N}\in \mathcal{N}\text{ and }\hat{C}\in \mathcal{C}_{\hat{N}}\}.
\]

%
\item join node: let $t_1$ and $t_2$ be the children of $t$. Recall that $X_{t_1}=X_{t_2}$. Then:
\[
T_t(N,C)=-\lvert N\rvert + \max\{T_{t_1}(N,C_1) + T_{t_2}(N,C_2)\mid C_1\cap C_2 = V^{T}(N)\text{ and }C_1\cup C_2 = C\}.
\]
\end{itemize}


\begin{theorem}
\label{TtMcorrect}
\TMatching{} can be computed in time $O^*(2^{w^2}\cdot 8^{w\cdot\tau})$.
\end{theorem}
\begin{proof}
We first prove that each entry of table $T$ can be computed with the presented recursive function.
So, consider $t\in V(T)$ and sets $N\subseteq E(X_t)$ and $C\subseteq V^T(X_t)$. We analyse each possible type of node $t$.

\begin{itemize}
    \item  $t$ is a leaf: then we can only have $N = \emptyset = C$, and there is only one subset $N'$ of $E(G_t)$, which is again the empty set and has cardinality $0$. Therefore $T_t(\emptyset,\emptyset)=0$.

    \item $t$ is an introduce node: let $v\in V(T)$ be introduced by $t$ and let $t'$ be its child. 
    Recall that, since $(T,\{X_t\}_{t\in V(T)})$ is a nice tree decomposition, each vertex of $G$ is forgotten precisely once, and for each edge $e$ of $G$ there exists an node in the tree that contains both endpoints of $e$. Therefore all the edges in $E(G_t)$ that are incident to $v$ are also contained in $E(X_t)$. This means that $F$ must saturate all the temporal vertices in $\{v\}\times[\tau]$ contained in $C$. Therefore if $V^{T}(F)\cap (\{v\}\times [\tau]) \neq C\cap (\{v\}\times [\tau])$, then there cannot be any temporal matching satisfying the properties required by $T_t(N,C)$ and, by definition, we get $T_t(N,C)=0$. 
    If instead $F$ saturates exactly the temporal vertices in $\{v\}\times[\tau]$ that are also in $C$, then we need to prove that
    \[
    T_t(N,C) = T_{t'}(N\setminus F, C\setminus V^T(F)) + |F|.
    \]
    To do this, we start with a matching in $G_t$ ($G_{t'}$) and construct a matching in $G_{t'}$ ($G_t$) by removing (adding) $F$. 
    So first let $M$ be a temporal matching of $G_{t}$ such that $M\cap E(X_t) = N$, $V^T(M)\cap V^T(X_t) = C$, and $T_t(N,C)\ge \lvert M\rvert$.
    Then $M\setminus F$ is a temporal matching on $G_{t'}$ that uses the edges $N\setminus F$ and saturates $C\setminus V^{T}(F)$, and $|M\setminus F|=|M|-|F|$.
    Therefore
    \[
    T_t(N,C) \le T_{t'}(N\setminus F, C\setminus V^T(F)) + |F|.
    \]
    To prove the other inequality we reason in the same way, this time adding $F$ to the temporal matching.

    \item $t$ is a forget node: let $v\in V(T)$ be forgotten by $t$ and let $t'$ be its child. Also, define $\mathcal{N}$ and $\mathcal{C}_{\hat{N}}$ as in the recursive definition. First let $M\in E(G_t)$ be such that $M\cap E(X_t) = N$, $V^T(M)\cap V^T(X_t) = C$, and $T_t(N,C)=\lvert M\rvert$. Define $\hat{N} = \delta_G(v)\cap E(X_{t'}) \cap M$ and $\hat{C} = V^T(M)\cap (\{v\}\times [\tau])$. Since $C$ is exactly the set of temporal vertices in $V^T(X_t)$ saturated by $M$ and $X_t = X_{t'}\setminus \{v\}$, we get that the endpoints of $\hat{N}$ distinct from $v$ must also be in $C$; hence, $\hat{N}\in \mathcal{N}$. Additionally, $\hat{C}$ must contain all the temporal vertices $(v,i)$ saturated by $M$ and, in particular, those that are in $V^T(\hat{N})$; hence $\hat{C}\in \mathcal{C}_{\hat{N}}$. By construction of $\hat{N}$  and $\mathcal{C}_{\hat{N}}$, observe that $M$ is a matching in $G_{t'} = G_t$ such that $M\cap E(X_{t'}) = N\cup \hat{N}$ and $V^T(M)\cap V^T(X_{t'}) = C\cup\hat{C}$. Therefore, $T_{t'}(N\cup \hat{N},C\cup\hat{C})\ge \lvert M\rvert = T_t(N,C)$, which means that the maximum taken over $\hat{N}$ in our recursive formula is also at least $T_t(N,C)$.
    On the other hand, one can pick a matching of $G_{t'} = G_t$ satisfying the conditions of the recursive function and observe that it also define a matching satisfying the conditions in the definition of $T_t(N,C)$, giving the opposite inequality.

    \item $t$ is a join node: let $t_1$ and $t_2$ be the children of $t$. First, consider a matching $M$ of $G_t$ such that $M\cap E(X_t) = N$, $V^T(M)\cap V^T(X_t) = C$, and $T_t(N,C) = \lvert M\rvert$.
    Let $i\in \{1,2\}$, and define $M_i = M\cap E(G_{t_i})$ and $C_i = V^T(M_i)\cap V^T(X_{t_i})$. Observe that, since $E(X_t) = E(X_{t_i})\subseteq E(G_{t_i})$, we get that $V^T(M_i)\cap E(X_{t_i}) = V^T(M)\cap E(X_t) = N$. Additionally, every temporal vertex saturated by an edge in $N$ is also saturated by $M_i$, giving $V^T(N)\subseteq C_i$. Finally, every temporal vertex saturated by $M$ must be saturated either by $M_1$ or by $M_2$ (or both), giving us $C_1\cup C_2 = C$. Observe that $T_{t_i}(N,C_i)\ge \lvert M_i\rvert$ by definition of $C_i$.
    Therefore $T_t(N,C) = \lvert M\rvert = \lvert M_1\rvert +\lvert M_2\rvert -\lvert N\rvert \le T_{t_1}(N,C_1)+T_{t_2}(N,C_2) - \lvert N\rvert$, which implies that the maximum taken over $C_1$ and $C_2$ in our recursive formula is also at least $T_t(N,C)$. 
    On the other hand, one can verify that picking temporal matchings $M_1$ and $M_2$ satisfying the conditions of the recursive function and letting $M = M_1\cup M_2$, we obtain a temporal matching satisfying the definition of $T_t(N,C)$.
\end{itemize}

Now, we analyse the running time needed to compute $T$. 
As already noticed in the proof of Theorem~\ref{thm:FPTEdgeCover_twplustau}, we know that there exists a nice tree decomposition with $O(n)$ nodes, where $n = \lvert V(G)\rvert$. Additionally, 
for each node $t\in V(T)$, there are at most $2^{\binom{|X_t|}{2}}$ subsets $N\subseteq E(X_t)$ and $2^{|V^T(X_t)|}$ subsets $C\subseteq V^T(X_t)$. We therefore have a table of size $O^*(2^{w^2 +w\cdot\tau})$. It remains to analyse the time needed to compute each entry of such table. For this, let $t\in V(T)$, $N\subseteq E(X_t)$, and $C\subseteq V^T(X_t)$. We consider each possible type of node $t$.

\begin{itemize}
    \item $t$ is a leaf: then  $T_t(\emptyset,\emptyset)=0$ and this is computed in time $O(1)$.

    \item $t$ is an introduce node: then checking whether $V^{T}(F)\cap (\{v\}\times [\tau]) = C\cap (\{v\}\times [\tau])$, where $F=N\cap \delta_{G}(v)$, takes time $O(w\cdot\tau)$. Since we assume to have already computed each value of $T_{t'}$, the computation of $T_{t}(N,C)$ for an introduce node takes time $O^*(w)$.

    \item $t$ is a forget node: then there are at most $2^{w}$ subsets in $\mathcal{N}$ and at most $2^\tau$ subsets in $\mathcal{C}$ and checking whether or not some set belongs to $\mathcal{N}$ or $\mathcal{C}$ takes time $O(w^2)$. Indeed, one needs to check whether the endpoints of $\hat{N}$ are within $C$ or $\hat{C}$; since each of these sets has size $O(w)$, our claim follows. We therefore get a total time of $O(w\cdot 2^{w+\tau})$.

    \item $t$ is a join node: then the number of combinations for $C_1$ and $C_2$ is at most $O(2^{2 w \cdot \tau})$.
\end{itemize}

Of these nodes, the worst case is the one for the join node. Hence each entry can be computed in time  $O(4^{w \cdot \tau})$, and the theorem follows.
\end{proof}

\section{Approximation of \TMatching{}}
\label{sec:ApproxMatching}

In this section we consider the approximability of \TMatching{}. We start by discussing a bound on the approximability of the problem.
Since the reduction described in Section~\ref{sec:HardMatchTree} is also approximation preserving (note that it defines $\tau = n$) and since \PSets{} is hard to approximated within factor $O(n^{1 - \varepsilon})$~\cite{DBLP:conf/coco/Karp72,DBLP:journals/toc/Zuckerman07}, for any $\varepsilon > 0$, unless P = NP, then we have the following result.

\begin{corollary}
\label{matchingtHardApprox}
\TMatching{} cannot be approximated within factor $O(\tau^{1-\varepsilon})$,
for any $\varepsilon > 0$, unless P = NP.    
\end{corollary}
\begin{proof}
In Section \ref{sec:HardMatchTree} we have designed an approximation preserving reduction from \PSets{} to \TMatching{}.
Since \PSets{} is hard to approximate within factor
$O(n^{1 - \varepsilon})$, for any $\varepsilon > 0$, unless P = NP~\cite{DBLP:conf/coco/Karp72,DBLP:journals/toc/Zuckerman07},
and $\tau = n$, it follows that
\TMatching{} cannot be approximated within factor $O(\tau^{1-\varepsilon})$,
for any $\varepsilon > 0$, unless P = NP.
\end{proof}

On the positive side, we can prove that 
\TMatching{} can be easily approximated within
factor $\tau$, by computing a maximum matching
in each snapshot and returning as approximated solution the one having maximum cardinality.

\begin{theorem}
\TMatching{} can be approximated in polynomial time
within factor $\tau$.
\end{theorem}
\begin{proof}
Consider the following approximation algorithm.
For each $t \in [\tau]$, the approximation algorithm computes a maximum matching $M_t$ of the static graph $G_t$, defined as $\mathcal{G}$ restricted to time $t$ (i.e. $G_t$ is the snapshot of $\mathcal{G}$ in $t$).
Then the approximation algorithm returns as an approximated solution, denoted by $M$, a matching of maximum cardinality among $M_t$, $t \in [\tau]$.

First, note that $M$ is a feasible solution of
\TMatching{}. Indeed, since $M$
is a matching in a static graph,
each pair of edges in $M$ is vertex disjoint, hence $M$ is also a temporal matching.
Now, we prove that the approximation factor is indeed
$\tau$.
Consider a maximum temporal matching $M^*$ in $\mathcal{G}$.
Consider the set of edges  
$M^*_t \subseteq M^*$ defined at time $t$, $t \in [\tau]$.
By definition of temporal matching, the 
edges in $M^*_t$ must be vertex disjoint,
thus they must be a matching in $G_t$. Since 
for each $t \in [\tau]$ $M_t$ is a maximum matching of $G_t$, it follows that $|M^*_t| \leq |M_t|$.
By construction of $M$, we have
\[
\sum_{t \in [\tau]} |M^*_t| \leq \sum_{t \in [\tau]} |M_t| \leq \tau |M|,
\]
thus concluding the proof.
\end{proof}

\section{Relation between Max Temporal Matching and Min Temporal Edge Cover}
\label{sec:rel}
In this section, we show that having a minimal temporal edge cover does not facilitate the computation of a maximum temporal matching, and vice versa.

For a static graph, the problem of finding the maximum size of a matching and the problem of finding the minimum size of an edge cover are complementary. More specifically, given a graph $G$ on $n$ vertices and denoting the size of a minimum edge cover by $\beta'(G)$ and the size of maximum matching by $\alpha'(G)$, it is known that $\alpha'(G)+\beta'(G) = n$. 
Indeed, we can even construct a matching from an edge cover, and vice-versa. 
To see this, let $M$ be a matching of size $k$. Picking $M$ plus one edge incident to each non-saturated vertex gives us an edge cover of size $k+n-2k$, thus implying that $\beta'(G)\le n-\alpha'(G)$. On the other hand, if $N$ is a minimal edge cover of cardinality $k$, observe that $G'=(V(G),N)$ is a forest of stars. Indeed, $G'$ contains no cycles as removing an edge of a cycle in $G'$ would cover the same vertices. Additionally, if $G'$ contains a path $P = (v_1,v_2,v_3,v_4)$, then $N-v_2v_3$ still covers $V(G)$. Let $k'$ be the number of components of $G'$ and observe that we can construct a matching of size $k'$ by picking one edge of each star of $G'$. Finally, it is known that a forest on $n$ vertices and $k'$ components has exactly $n-k'$ edges, i.e., $k = n-k'$, from which we get $\alpha'(G)\ge n-\beta'(G)$. 

We now see that the temporal variants of matchings and edge covers are not related as in the static case. 
That is, given a temporal graph $\mathcal{G}$ and a temporal matching of maximum cardinality, the problem of finding a minimum temporal edge cover for $\mathcal{G}$ is still NP-complete. The opposite is also true, which means that if we are given a minimum temporal edge cover then the problem of finding a maximum temporal matching is still NP-complete. To see this, we use some of the reductions presented throughout the paper.

Let $S_1$, \dots, $S_m$ be an instance of \SetCover{}. \cref{NPtree} and \cref{treeedgecover} detail a reduction to \TEdgeCover{}, where the resulting temporal graph $\mathcal{G}=(G,\lambda)$ has lifetime $\tau=\max\{k\mid k\in S_i, 1\le i\le m\}$. We now construct a temporal graph $\hat{\mathcal{G}}=(G,\mu)$ with lifetime $\tau+1$ where $\mu(e)=\lambda(e)\cup\{\tau+1\}$, for each $e\in E(G)$. That is, we add $\tau+1$ to each label. Then any temporal matching of maximum cardinality for $\hat{\mathcal{G}}$ contains all the edges $x_i y_i$, $1\le i\le m$ for each $i$ except for at most one $j$, and in that case it contains $r x_j$. This does not depend on the specific instance $S_1$, \dots, $S_m$ considered. Still, any temporal edge cover of minimum cardinality is a solution for our instance of \SetCover{}, since the addition of the same element $\tau+1$ to all labels does not change which edges are a solution.
Therefore having a temporal matching of maximum cardinality does not change the complexity of finding a temporal edge cover of minimum cardinality.

On the other hand, suppose that for a temporal graph we know all its temporal edge covers of minimum cardinality, and we want to find a temporal matching of maximum cardinality. Then we can use the reduction from packing set detailed in \cref{matchingtreeNP} and \cref{TreeMatchingimage}. Indeed, the only edge cover takes all the edges of the graph, but the matching depends on the specific sets $S_1$, \dots, $S_n$. Thus having a temporal edge cover of minimum cardinality does not change the complexity of finding a temporal matching of maximum cardinality.

\section{Conclusion}

In this paper, we have investigated the computational complexity of \textsc{Edge Cover} in temporal graphs. We quickly identified the most interesting case (see again \Cref{tab:ECvariants}), which we simply named \textsc{Temporal Edge Cover}. We presented two NP-completeness results for this problem, one which uses lifetime $\tau = 2$, and another where the underlying graph is a tree (i.e. treewidth equals $1$). These results complement our following FPT result, as the parameters considered in our proposed algorithm are $\tau$ and treewidth. Then, we have explored approximation of \TEdgeCover{} and provided an approximation algorithm with an asymptotically tight approximation factor of $O(\log \tau)$. Inspired by the intrinsic connection between \textsc{Edge Cover} and \textsc{Matching} in (non-temporal) graphs, we also have provided such results for \textsc{Temporal Matching}. Surprisingly, even though the problems are shown to be distinct and unrelated to each other in the temporal setting, we have proved very similar results for both (albeit through different reductions and observations).\\
Although we have presented a comprehensive overview, covering classical complexity, parameterized complexity in terms of lifetime and treewidth, and approximation, we identify the following directions for future research. It may be interesting to identify specific classes of temporal graphs for which tractability of the (non-parametrised) problems is possible, and even more so if these classes correspond to a natural setting for real-life applications of our problems (e.g. \TEdgeCover{} in planar graphs possibly representing surveillance of a building floor). In terms of parametrized complexity, other parameters can be considered, such as some recently introduced parameters specifically for temporal graphs (see, e.g., the parameters studied and mentioned in \cite{enright2024structural}).



\end{document}